\documentclass[lettersize,journal]{IEEEtran}
\usepackage{amsmath,amsfonts}
\usepackage{algorithmic}
\usepackage{array}
\usepackage[caption=false,font=normalsize,labelfont=sf,textfont=sf]{subfig}
\usepackage{textcomp}
\usepackage{stfloats}
\usepackage{url}
\usepackage{verbatim}
\usepackage{graphicx}
\usepackage{enumerate}
\usepackage[T1]{fontenc}
\usepackage{multirow}
\usepackage{makecell}
\usepackage{threeparttable}    
\usepackage[table]{xcolor}
\usepackage{listings}
\usepackage{mdframed}
\usepackage{lineno} 
\usepackage{xcolor}
\usepackage{letterspace}
\usepackage{float}

\lstdefinelanguage{Solidity}{
	keywords=[1]{anonymous, assembly, assert, balance, break, call, callcode, case, catch, class, constant, continue, constructor, contract, debugger, default, delegatecall, delete, do, else, emit, event, experimental, export, external, false, finally, for, function, gas, if, implements, import, in, indexed, instanceof, interface, internal, is, length, library, log0, log1, log2, log3, log4, memory, modifier, new, payable, pragma, private, protected, public, pure, push, require, return, returns, revert, selfdestruct, send, solidity, storage, struct, suicide, super, switch, then, this, throw, transfer, true, try, typeof, using, value, view, while, with, addmod, ecrecover, keccak256, mulmod, ripemd160, sha256, sha3}, 
	keywordstyle=[1]\color{blue}\bfseries,
	keywords=[2]{address, bool, byte, bytes, bytes1, bytes2, bytes3, bytes4, bytes5, bytes6, bytes7, bytes8, bytes9, bytes10, bytes11, bytes12, bytes13, bytes14, bytes15, bytes16, bytes17, bytes18, bytes19, bytes20, bytes21, bytes22, bytes23, bytes24, bytes25, bytes26, bytes27, bytes28, bytes29, bytes30, bytes31, bytes32, enum, int, int8, int16, int24, int32, int40, int48, int56, int64, int72, int80, int88, int96, int104, int112, int120, int128, int136, int144, int152, int160, int168, int176, int184, int192, int200, int208, int216, int224, int232, int240, int248, int256, mapping, string, uint, uint8, uint16, uint24, uint32, uint40, uint48, uint56, uint64, uint72, uint80, uint88, uint96, uint104, uint112, uint120, uint128, uint136, uint144, uint152, uint160, uint168, uint176, uint184, uint192, uint200, uint208, uint216, uint224, uint232, uint240, uint248, uint256, var, void, ether, finney, szabo, wei, days, hours, minutes, seconds, weeks, years},	
	keywordstyle=[2]\color{teal}\bfseries,
	keywords=[3]{block, blockhash, coinbase, difficulty, gaslimit, number, timestamp, msg, data, gas, sender, sig, value, now, tx, gasprice, origin},	
	keywordstyle=[3]\color{violet}\bfseries,
	identifierstyle=\color{black},
	sensitive=true,
	comment=[l]{//},
	morecomment=[s]{/*}{*/},
	commentstyle=\color{gray}\ttfamily,
	stringstyle=\color{red}\ttfamily,
	morestring=[b]',
	morestring=[b]"
}

\def\BibTeX{{\rm B\kern-.05em{\sc i\kern-.025em b}\kern-.08em
    T\kern-.1667em\lower.7ex\hbox{E}\kern-.125emX}}
\usepackage{balance}
\begin{document}

\title{A Comparative Evaluation of Automated Analysis Tools for Solidity Smart Contracts}
\author{Zhiyuan Wei, Xianhao Zhang, Jing Sun, Zijian Zhang, Zhen Li, Meng Li, Yuqiang Sun, Yue Xue, Yang Liu, Liehuang Zhu
\thanks{Manuscript received October 1, 2023. \textit{(Corresponding author: Zijian Zhang, Meng Li)}.

Xianhao Zhang, Zijian Zhang and Liehuang Zhu are with the School of Cyberspace Science and Technology at Beijing Institute of Technology, Beijing, China (e-mail: zhangxh@bit.edu.cn; zhangzijian@bit.edu.cn; liehuangz@bit.edu.cn).

Zhiyuan Wei and Zhen Li with the School of Computer Science and Technology at Beijing Institute of Technology, Beijing, China (e-mail: weizhiyuan@bit.edu.cn; zhen.li@bit.edu.cn).

Jing Sun is with the School of Computer Science at University of Auckland, New Zealand (e-mail: jing.sun@auckland.ac.nz).

Meng Li is with the School of Computer Science and Information Engineering at Hefei University of Technology, Hefei, China (e-mail: mengli@hfut.edu.cn).

Yuqiang Sun and Yang Liu are with the School of Computer Science and Engineering at Nanyang Technological University, Singapore, Singapore (e-mail: suny0056@e.ntu.edu.sg; yangliu@ntu.edu.sg).

Yue Xue is with MetaTrust Labs, Singapore, Singapore (e-mail: xueyue@metatrust.io).
}}

\markboth{Journal of \LaTeX\ Class Files,~Vol.~18, No.~9, August~2023}%
{How to Use the IEEEtran \LaTeX \ Templates}

\IEEEpubid{0000--0000/00\$00.00~\copyright~2021 IEEE}

\maketitle

\begin{abstract}
Since Solidity smart contracts manage trillions of dollars in virtual assets across various domains, they have become prime targets for cyberattacks. A significant challenge for the academic community is to objectively and comprehensively evaluate and compare existing automated tools for detecting vulnerabilities in smart contracts. However, current studies face limitations in three key areas: evaluation criteria, timeliness, and decision-making.

In this paper, we aim to address these limitations in four key areas:
(1) We propose a novel criterion for evaluating these tools, based on the ISO/IEC 25010 standard, one of the most recent international standards for software quality assessment.
(2) Using these evaluation criteria, we develop a scoring framework for smart contract analysis tools by leveraging two decision-making algorithms.
(3) We construct a benchmark that encompasses two distinct datasets: a collection of 389 labeled smart contracts and a scaled set of 20,000 unique cases from real-world contracts. The first dataset allows us to assess how well the tools handle a variety of specific vulnerabilities, while the second provides a more realistic test of their performance on practical smart contracts.
(4) We examine 13 automatic tools, focusing specifically on their ability to identify vulnerabilities in Solidity smart contracts and analyzing the differences introduced by various vulnerability detection techniques.

The entire evaluation process involves 85,057 scanning tasks (13 × 389 + 4 × 20,000), empirically validating the effectiveness of the scoring framework and offering a new approach for evaluating smart contract analysis tools. Through this evaluation, we aim to provide developers and researchers with valuable guidance on selecting and using smart contract analysis tools and contribute to ongoing efforts to improve the security and reliability of smart contracts. Our evaluation results, along with benchmark datasets, are available at \url{https://github.com/bit-smartcontract-analysis/smartcontract-benchmark}.

\end{abstract}


\begin{figure*}[ht]
  \centering
  \includegraphics[width=4.5in]{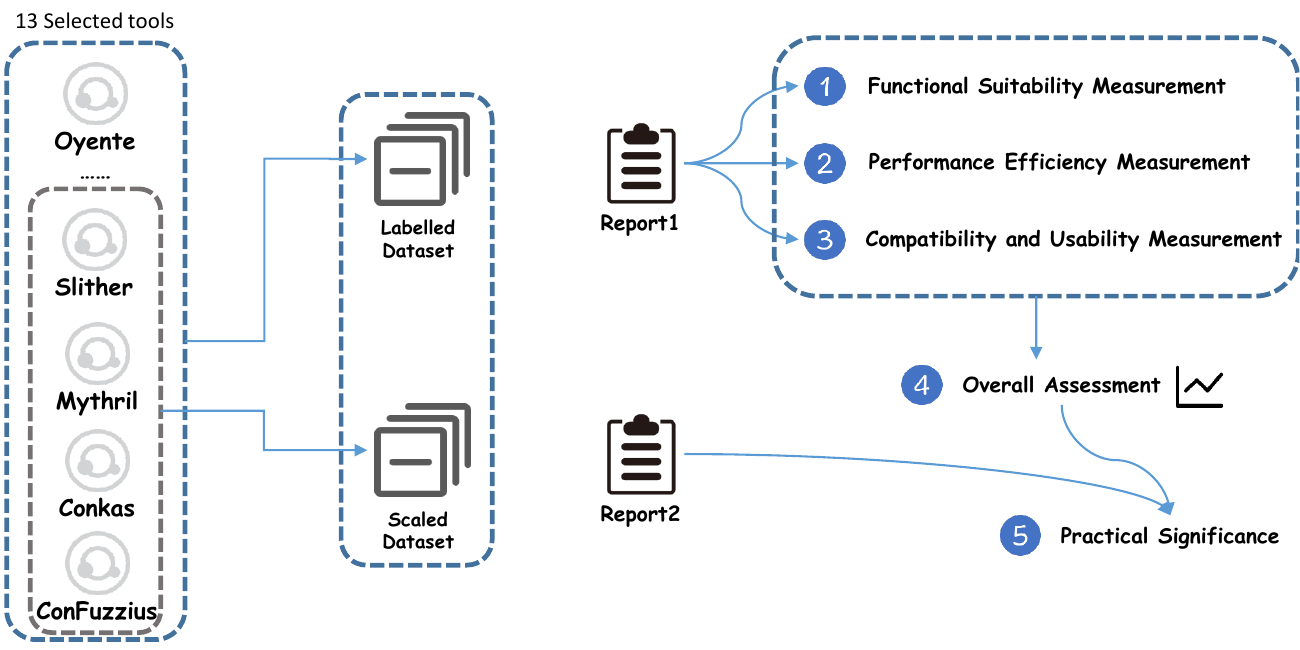}
  \caption{Overview of our study}
  \label{overview}
\end{figure*}

\section{Introduction}
\IEEEPARstart{S}{mart} contracts, defined as self-executing agreements encoded in software, have gained popularity across diverse domains where security and privacy are crucial \cite{Luu16}. They enable transactions and agreements to be carried out without the need for a central authority mechanism or system by a distributed, decentralized blockchain network. And owing to the immutable nature of the blockchain, any smart contract deployed therein strictly adheres to its predefined rules during execution. 
\IEEEpubidadjcol
Recently, the majority of blockchain platforms have incorporated support for smart contracts. Ethereum \cite{Ethereum}, as a pioneering blockchain platform, expanded the range of blockchain applications and use cases through the introduction of smart contracts. As a consequence, it expanded the range of possible applications and use cases for blockchain technology. Prior to Ethereum's advent, the blockchain was primarily associated with cryptocurrency transactions. However, with the introduction of smart contracts, platforms such as Ethereum, Hyperledger Fabric \cite{HF}, EOSIO \cite{eosio}, and ChainMaker \cite{chainmaker} have opened up avenues for more complex, programmable transactions and applications. These applications go beyond simple asset transfers, including Decentralized Finance (DeFi), supply chain management, voting systems, and Decentralized Autonomous Organizations (DAOs).

However, solidity smart contracts have seen numerous instances of high-profile vulnerabilities and exploits in the past \cite{Cui2022, Bose2022}. 
One infamous case occurred in 2016, when attackers exploited vulnerabilities to divert more than 3.6 million ether into a separate entity known as a child DAO \cite{Falkon2017}. Another notable incident involved a significant vulnerability in the Parity Multisig Wallet, resulting in a financial loss approximating 30 million USD \cite{Palladino2017}.
Secure contracts is crucial for their development and wider application. Furthermore, creating secure smart contracts are far from a trivial task. In a preliminary study performed on 19,366 Ethereum contracts, 8,833 of them were flagged as vulnerable, accounting for nearly half of the contracts examined \cite{AtzeiBC16}. Another study by Nikolić et al. underscores these findings \cite{Nikoli2018}, discovering that 34,200 out of nearly one million Ethereum contracts analyzed were flagged as vulnerable. These studies suggest that the current state of smart contract development might not be adequately addressing security concerns, highlighting the importance of robust smart contract analysis and stringent security practices.
Comprehensive analysis allows developers to proactively identify and mitigate potential risks prior to deploying smart contracts in a production environment. Manual examination of smart contracts, while possible, can be time-consuming and prone to errors. Consequently, the development of automatic analysis tools, capable of efficiently detecting vulnerabilities, has gained importance. 

Automatic analysis tools leverage a variety of techniques to automate and enhance the efficiency of the analysis process \cite{AtzeiBC18}. Firstly, these tools can process extensive codebases, libraries, and frameworks, thereby conserving developers' time and effort. Secondly, they can conduct exhaustive scans of smart contract code to detect a multitude of vulnerabilities and potential issues. Lastly, they can minimize the risk of human error and ensure consistent, reliable results. The Ethereum research community has actively contributed to the development of automated analysis tools \cite{Chen2021}, aiming to enhance the security and reliability of smart contracts.

Previous researchers have made pioneering efforts in evaluating automated vulnerability detection tools for smart contracts, such as constructing annotated benchmarks, developing standardized tool execution frameworks, and conducting in-depth analysis of detection effectiveness for specific vulnerabilities based on experimental results. However, we still find that existing work often exhibits the following limitations:
\begin{enumerate}
    \item Evaluation Metric Limitations: Past researches \cite{Durieux2020, ghaleb2020effective, ren2021empirical} have primarily focused on metrics such as Accuracy, Precision, Recall, and F1-score, along with evaluating the time performance of tools. This results in relatively narrow evaluation dimensions. A recent study \cite{li2024static} has started to consider issues like vulnerability coverage and compatible contract versions for tools, but the analysis of these metrics remains fragmented, lacking a structured, unified standard to guide these evaluations.
    \item Temporal Limitations: This limitation manifests in two ways. First, smart contract vulnerability detection tools are continuously evolving, so previous evaluation results may not apply to the latest tool versions. Second, contract versions and vulnerability patterns are also constantly changing, introducing new challenges for the design of analysis tools. Traditional evaluations tend to focus more on the short-term performance of tools, with insufficient consideration of their applicability and maintainability in various environments.
    \item Decision-Making Limitations: Beyond detection capabilities, factors such as usability, maintainability, and compatibility are also essential for an overall and objective evaluation of these tools. A study \cite{wei2024survey} has attempted to assign weights to each metric and simply calculate a composite score by summing the weighted metrics. However, this weighting approach is overly subjective, and the metrics are not normalized, resulting in scores that lack sufficient persuasiveness.
\end{enumerate}

In this paper, we aim to conduct a comprehensive review of the state-of-the-art automated analysis tools for smart contracts to address the aforementioned limitations as well as to provide recommendations. The overview of our study is illustrated in Figure \ref{overview}, and our key contributions can be summarized as follows:

\begin{itemize}
\item{\textbf{Evaluation Criteria}}: We formulated a unique set of evaluation criteria, grounded in the principles of ISO/IEC 25010 \cite{ISO25010}, for assessing the quality of smart contracts. This set of criteria provides a structured approach for systematically evaluating software product quality, thereby offering a robust means of comparing different smart contract analysis tools.
\item{\textbf{Scoring Framework}}: Based on the evaluation criteria, we developed a scoring framework for smart contract analysis tools. This framework utilizes the entropy weight method and the analytic hierarchy process from the Multi-Criteria Decision Making domain to determine indicator weights. After normalizing the indicators, the final scores are calculated. To the best of our knowledge, this is the first scoring framework specifically designed for smart contract analysis tools, offering a novel approach for comprehensive, objective, and effective evaluation of such tools.
\item{\textbf{Benchmarking Dataset}}: We assembled a benchmark that comprises two distinct datasets: a collection of 389 labelled smart contracts across a diverse array of vulnerability types, and a large set of 20,000 real-world contract cases. This labelled dataset serves as a concentrated evaluation ground to test the tools' efficacy across specific, well-defined scenarios. This large dataset aims to offer a wide-angle view, enabling evaluations that reflect the practical complexities and evolving trends in smart contract deployments. Both datasets are publicly accessible and can be found at \url{https://github.com/bit-smartcontract-analysis/smartcontract-benchmark}.
\item{\textbf{Technical Coverage}}: We conducted an in-depth evaluation of 10 representative smart contract analysis tools across a range of five major automatic analysis techniques, including formal verification, symbolic execution, fuzzing, intermediate representation, and machine learning. Our evaluation offers a comparative study of these tools based on the proposed set of evaluation criteria, spotlighting their individual strengths and shortcomings.
\end{itemize}

The rest of this paper is structured as follows. Section \ref{background} presents the current challenges in the field of smart contracts and automatic analysis tools. In Section \ref{methodology}, we explore our research methodology, which includes the tool section and evaluation criteria. In Section \ref{evaluation}, we create our datasets and perform the evaluation on selected tools. In Section \ref{discussion}, we present our findings and discuss threats to validity. Finally, in Section \ref{conclusion}, we conclude by summarizing this paper and offering insights into the future trajectory of smart contract analysis tools in light of the identified challenges and suggested solutions.

\section{Backgrounds}
In this section, we introduce background knowledge of our work, including smart contracts, automatic analysis methodologies for smart contracts, and ISO/IEC 25010, which is one of the international standards of software quality.

\label{background}
\subsection{Smart Contracts}
\begin{figure}[tp]
  \centering
  \includegraphics[width=3.5in]{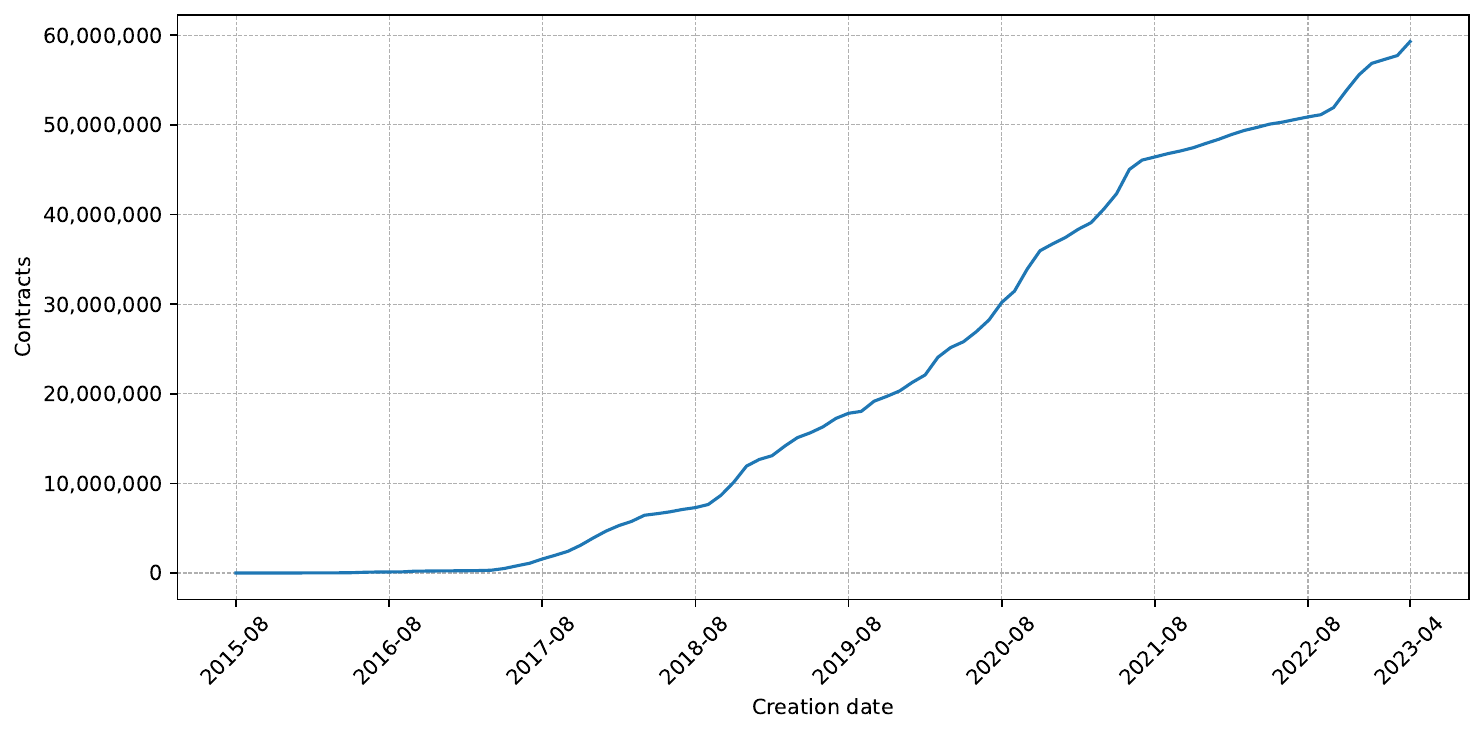}
  \caption{Number of smart contracts over time}
  \label{fig_contracts_creation}
\end{figure}

Smart contracts are designed to digitally facilitate, verify, and enforce the performance of a contract, all while avoiding the need for a third party. They not only define the rules and penalties related to an agreement in a similar way to a traditional contract but also automatically enforce these obligations.
The term \textit{smart contract} was first coined by computer scientist Nick Szabo in 1994, long before the advent of blockchain technology \cite{Szabo1996}. Szabo posited that a digital contract could be made smart through programmable scripts that execute contractual clauses automatically when predefined conditions are met. The introduction of Bitcoin laid the groundwork for blockchain technology in 2009, bringing the underlying technology of smart contracts \cite{Nakamoto2008}. However, until Ethereum officially launched in 2015, the Turing-complete smart contracts became a reality with its own programming language, Solidity. Figure \ref{fig_contracts_creation} demonstrates a significant growth in the number of contracts over the past eight years. This upward trend could reflect various factors, such as increased adoption, technological advancements, or broader market dynamics within the context of the data.
These smart contracts can be used to facilitate a wide range of applications, from the standardization of token contracts (such as ERC20 and ERC721 standards) to decentralized finance. As of now, smart contracts are being used in a variety of sectors beyond finance, including supply chain, healthcare, and more.

One important characteristic of Ethereum smart contracts is their immutability after deployment. This means that once a smart contract is established on the Ethereum network, it cannot be modified, thus leading to a strong and irreversible commitment to a specific functionality. This attribute is part of what makes blockchain technology so secure and trustworthy, i.e., transactions or agreements encoded in smart contracts cannot be tampered with or retroactively changed. However, this feature of immutability also presents its own unique set of challenges. Since smart contracts cannot be altered once they are deployed, any existing bugs, flaws, or vulnerabilities in the contract's code become permanently embedded within the contract. Unlike traditional software where patches or updates can be issued to correct bugs or flaws, smart contracts offer no such flexibility. This means that if a vulnerability is discovered after deployment, it cannot be rectified in the affected contract. Such vulnerabilities, if exploited by malicious actors, can lead to substantial financial losses. A prime example of this is the DAO hack in 2016 \cite{Falkon2017}. A flaw in the DAO's smart contract was exploited, leading to the theft of over \$50 million worth of Ether. Furthermore, the nature of blockchain networks allows smart contracts to interact with other contracts and transactions in complex ways. This complexity makes it challenging to predict all possible interactions and outcomes.

\subsection{Automatic Analysis Methodology}

As blockchain technology continues to grow and diversify, there is an increasing reliance on the quality and integrity of smart contracts. When Ethereum was launched in 2015, the ecosystem for smart contract development was in its infancy, with few tools available for testing and verification. Developers primarily relied on manual testing and code reviews to verify the correctness and security of their contracts. However, manual examination of smart contracts is often labour-intensive and prone to human error. As the complexity and deployment of smart contracts have increased, there has been a corresponding surge in the demand for analysis methods.
To address these challenges, both the industry and the research community have been actively working to enhance the methods for analyzing and evaluating smart contracts.
These tools are designed to systematically identify vulnerabilities, bad practices, and other potential issues within the smart contract implementation. It allows developers to proactively address problems before the contract is deployed, significantly reducing the risk of exploitation. These automated analysis tools employ various techniques, such as formal verification, symbolic execution, fuzzing, intermediate representation, and machine learning, which are detailed as follows:

\begin{itemize}
  \item \textbf{Formal verification} is a mathematical-based technique to build and check formal proofs that satisfy a particular property \cite{Tolmach2022}. It is applied to ensure that software behaves and performs as expected in its specifications and requirements based on large reachable state spaces. In the context of smart contracts, formal verification can be used to prove that a contract behaves exactly as specified under all possible conditions.  It can provide a very high level of assurance about a contract's correctness. It can also detect subtle vulnerabilities that might be missed by other types of analysis.
  \item \textbf{Symbolic execution} is a technique used to find deep program errors by exploring all possible execution paths \cite{Lin2022}. It doesn't require specified input values; instead, it abstracts them into symbols. However, with large and complex programs, symbolic execution can generate too many paths, leading to a problem known as \textit{path explosion}. Fortunately, symbolic execution remains a popular choice for finding bugs and security vulnerabilities in smart contracts, given their maximum size is typically capped at 24KB.
  \item \textbf{Fuzzing} is a software-testing technique that executes target programs by inputting a large amount of abnormal or random test cases \cite{Li2021}. One advantage of fuzzing is that it can explore a large space of possible inputs and states, which can be hard to cover with manual testing. In the context of smart contracts, fuzzing is especially good at finding unexpected or undefined behaviour in a contract that might not be covered by manual tests. However, the quality of fuzzing largely depends on the quality and diversity of the input data. 
  \item \textbf{Intermediate Representation (IR)} is an abstraction of machine language used in compilers and programming language implementation \cite{Feist2019}. In the context of smart contract analysis, an Ethereum smart contract, which is originally written in a high-level language such as Solidity, is converted into an intermediate representation that is easier to analyze. This representation retains all the necessary details for analysis but simplifies the structure of the code to make the analysis more efficient and effective. The use of an intermediate representation can greatly increase the speed and efficiency of analysis tools, especially for complex smart contracts.
  \item \textbf{Machine learning} is a subfield of artificial intelligence that allows computers to learn from and make decisions or predictions based on data. In the context of smart contract analysis, machine learning techniques can be used to automatically identify patterns and detect anomalies or vulnerabilities in the contract's code. Machine learning can greatly enhance the capabilities of analysis tools, allowing them to detect more subtle or complex vulnerabilities and adapt to new types of vulnerabilities as they emerge. However, the effectiveness of machine learning techniques is heavily dependent on the quality and quantity of the training data. 
\end{itemize}

\subsection{ISO/IEC 25010}
ISO/IEC 25010 \cite{ISO25010, Cai2020} is an international standard that serves as a comprehensive framework for evaluating the quality of software products. As an extension of the ISO/IEC 9126 standard \cite{Jung2004}, ISO/IEC 25010 encompasses broader and more nuanced quality characteristics, positioning itself as a comprehensive benchmark in the landscape of software engineering.
ISO/IEC 25010 categorizes Product Quality Model properties into eight characteristics:

\begin{itemize}
    \item Functional suitability concerns whether the functions meet stated and implied needs, not the functional specification, including functional completeness, functional correctness, and functional appropriateness.
    \item Performance efficiency measures resource utilization under defined conditions, capturing aspects like time behavior and capacity.
    \item Compatibility evaluates the software's ability to exchange information with other systems and perform its required functions in different environments.
    \item Usability focuses on the ease with which specified users can achieve particular objectives effectively.
    \item Reliability considers the software's capability to perform its designated functions consistently over a given time period and under specified conditions.
    \item Security concerns whether a product protects information and data so that persons or other products have the degree of data access appropriate to their types and levels of authorization.
    \item Maintainability investigates the ease with which the software can be modified to adapt to environmental changes.
    \item Portability assesses the ease with which the software can be transferred from one computing environment to another.
\end{itemize}

These characteristics offer a flexible schema, allowing for tailored evaluation criteria depending on the application context. For example, in the domain of mobile computing, attributes like reliability and security may take on greater importance and should be assessed with criteria that are specially designed for mobile platforms \cite{Idri2013}.

\section{Methodology}
\label{methodology}

In this section, we present a systematic approach to evaluating the target analysis tools for smart contracts. This section is organized into three main parts: guiding research questions; the process of tool selection, including the criteria for selection, a description of the selected tools; and the establishment of our assessment criteria. By outlining our methods and criteria, we aim to provide a relatively objective, transparent, and easily reproducible framework for assessing both existing and future smart contract security analysis tools.

\subsection{Research Questions}
The foundation of our study lies in the following research questions that guided our investigation. These questions focus on various aspects of smart contract analysis tools.

\begin{enumerate}[RQ1:]
  \item \textbf{Functional Suitability Measurement. Does the analysis tools correctly analyze the smart contracts, providing accurate and reliable results?}
  This metric is the most important aspect in evaluating vulnerability detection tools. We list data on Accuracy, Precision, Recall, and F1-score for each tool across various types of vulnerabilities, aiming to provide researchers with a straightforward comparison.
  \item \textbf{Performance Efficiency Measurement. How is the performance of the target analysis tools in terms of detection time?} 
  Some types of vulnerabilities, such as Transaction Ordering Dependence and Bad Randomness(\ref{Vulnerability Types}), require high timeliness for exploitation. Thus, detecting vulnerabilities earlier can result in less financial loss, posing a significant challenge for the detection efficiency of these tools. However, due to limitations in detection technology, tools that utilize dynamic analysis techniques often require extensive analysis time, even for relatively small contracts.
  \item \textbf{Compatibility and Usability Measurement. To what extent do target analysis tools support different Solidity compiler version and vulnerability?} Tools with broad compatibility and ease of use can be adopted by a wider range of developers, enhancing the overall security landscape. Over time, the types of vulnerabilities in smart contracts continue to evolve, and Solidity compiler versions are constantly advancing. Some academic contract analysis tools face dual limitations of time and technology, making them unable to detect certain types of vulnerabilities or support the detection of higher-version contracts.
  \item \textbf{Overall Assessment. How to evaluate tools comprehensively, objectively, and effectively based on experimental data?}  To address this issue, we proposed a new scoring framework, employing two decision-making methods from operations research and management science to rate contract analysis tools: the entropy weight method and the analytic hierarchy process. Both are commonly used in the field of Multi-Criteria Decision Making (MCDM). The former automatically determines the weight of each evaluation criterion to provide a score, while the latter requires human quantification of the importance relationships between criteria, generating weights based on a consistency matrix to ultimately yield a score.
  \item \textbf{Practical Significance. What is the outcome of tools on real-world contracts?} Analyzing the outcome of tools on actual real-world contracts provides insights into how these tools function in practice. This helps in identifying potential gaps between theoretical capabilities and practical effectiveness, ensuring that the tools are robust and adaptable to real-world complexities.
\end{enumerate}

Our study aims to offer a comprehensive examination of the current state of smart contract analysis tools. Our goal is to provide insights that can inform both practitioners and researchers in the development and deployment of more secure and reliable smart contracts.

\subsection{Tools Selection}
The selection of appropriate analysis tools is a pivotal aspect of our study. The selected tools must be capable of addressing the specific research questions posed. A mismatch between the tools and the questions can lead to inconclusive or irrelevant findings. Thus, we define a set of selection criteria to help our study in filtering the available tools.

\subsubsection{Selection Criteria}
The selection criteria define the parameters used to choose the target analysis tools for the study. From an extensive compilation of 82 smart contract analysis tools gathered from academic literature and online resources, we applied a set of stringent selection criteria to identify the most suitable candidates for our evaluation. These criteria are designed to ensure that the selected tools align with the specific needs and goals of our study, as well as with the broader context of smart contract analysis. The key attributes of the compiled tools, including venue, methodology, input object, open-source link, and vulnerability ID, are detailed in the Appendices. The selection criteria applied are as follows:

\begin{enumerate}[C1:]
\item {[Availability]} The tool must be publicly accessible in an open-source manner for download or installation. It must also support a command-line interface (CLI) for integration into automated workflows and batch processing. This ensures that the tool is both accessible and adaptable, serving a wide range of users and use-cases without requiring further extensions.
\item {[Functionality]} The tool must be purpose-built for smart contract analysis, with a specific focus on detecting vulnerabilities. This criterion ensures that the tool's primary functionality aligns with the core objective of vulnerability detection, excluding constructs such as control flow graphs that are not directly involved in this process.
\item {[Compatibility]}  The tool must be compatible with the source code of smart contracts, written in Solidity. This excludes tools that solely focus on Ethereum Virtual Machine (EVM) bytecode, thereby ensuring that the tool is usable for source-level analysis. This facilitates more in-depth analyses that consider the high-level logic and semantics of the contract, rather than just the compiled bytecode.
\item {[Methodology \& Relevance]} The tool must employ at least one of the commonly recognized methods for smart contract analysis: formal verification, symbolic execution, fuzzing, static analysis(including IR) or machine learning. In our final selection of tools, we primarily focused on the number of citations the tools have received in research papers and the number of stars on their GitHub repositories. These metrics helped us identify tools with significant academic influence and practical relevance.
\end{enumerate}

\begin{table}[ht]
  \caption{Tool Selection Based Criteria}
  \centering
  \label{selected_tools}
  \begin{tabular}{p{1.8cm}<{\raggedright} | p{6cm}<{\raggedright}l}
  \hline
  \textbf{Criteria} & \textbf{Automatic Analysis Tool} \\
  \hline
  NOT Availability & ReDefender, EtherGIS, SmartMixModel, SolSEE, EXGen, Vulpedia, EOSIOAnalyzer, solgraph, VRust, CodeNet, eTainter, SmartFast, SolChecker, SciviK, SmartScan, SolGuard, Solidifier, ReDetect, GasChecker, ContractWard, ESCORT, VeriSolid, Ethainter, eThor, SmartSheild, GASOL, RA, EthPloit, Harvey, sCompile, FEther, NOChecker, SoliAudit, F framework, Zeus, SASC, Reguard, SVChecker \\
  \hline 
  NOT Functionality & solgraph, Vulpedia, VetSC, SciviK SmartPulse, Solythesis, Gastap, VeriSol, ContractLarva, EthIR, ESBMC, SIF \\
  \hline 
  NOT Compatibility & DefectChecker, EtherGIS, EthVer, Horus, Frontrunner-Jones, ContractGuard, VerX, SODA, TxSpector, ÆGIS, Sereum, Annotary, SolidityCheck, SolAnalyser, KEVM, Isabelle/Hol, EtherTrust, SolMet, Vandal, teEther, SaferSC, ContractFuzzer, MadMax, SKLEE, SVChecker, ILF \\
  \hline 
  NOT Methodology \& Relevance &  Echidna, Octopus, SODA, Solythesis, TxSpector, ÆGIS, Sereum, ECFChecker, Sailfish, Osirs\\
  \hline \hline 
  \textbf{Fully Compliant} & Oyente, Mythril, Slither, sFuzz, Conkas, Eth2Vec, GNNSCVD, ConFuzzius, VeriSmart, Securify, Solhint, Smartcheck, Maian 
  
  (HoneyBadger, Manticore) \\  
  \hline
  \end{tabular}
  \end{table}

\subsubsection{Tool Descriptions}
By rigorously applying our established selection criteria, we succeeded in narrowing down our extensive compilation of analysis tools to a refined set of 15 highly suitable automated analysis tools. 
However, Manticore \cite{MossbergMHGGFBD19} suffers from severe timeout issues, making it impossible to obtain accurate information about its performance. Additionally, the HoneyBadger \cite{torres2019art} specifically targets honeypot vulnerabilities in smart contracts, which fall outside the scope of this study. Therefore, we excluded these two tools and focused our evaluation on the remaining 13 tools.

The details of the selected tools, along with the tools that did not meet each specific criterion, are presented in Table \ref{selected_tools}. These selected tools represent a diverse and robust cross-section of the available methodologies and approaches to smart contract analysis.

For each of these techniques, we have selected two representative tools detailed as follows:

\begin{itemize}
    \item \textit{Securify} \cite{Securify}: Securify utilizes both compliance and violation patterns to symbolically encode the dependency graphs of the contracts. Semantic information is then extracted from EVM opcodes. This pattern-based rule system ensures a comprehensive and accurate vulnerability analysis.
    \item \textit{VeriSmart} \cite{SoLPLO20}:  VeriSmart specializes in arithmetic safety within Ethereum smart contracts. It employs formal verification methods that are grounded in mathematical models. This makes VeriSmart particularly adept at defending against arithmetic-related vulnerabilities, such as overflows and underflows. 
    \item \textit{Mythril} \cite{Mythril}: Developed by ConsenSys Diligence, Mythril is a multi-faceted tool that combines symbolic execution with taint analysis and control flow checking. Its versatility makes it a comprehensive tool for smart contract security. Furthermore, Mythril can be seamlessly integrated into various development tools and environments, adding to its utility.
    \item \textit{Oyente} \cite{AtzeiBC16}: As the first tool designed for automated smart contract analysis, Oyente has had a significant impact on the field. It employs symbolic execution to construct an accurate control-flow graph (CFG) of each compiled smart contract. And its pioneering role in the Ethereum ecosystem has influenced the development of subsequent smart contract analysis tools.
    \item \textit{ConFuzzius} \cite{TorresIGS21}: ConFuzzius is a unique tool that combines fuzzing with lightweight symbolic execution. It employs dynamic data dependency analysis to generate efficient transaction sequences. Additionally, ConFuzzius utilizes a constraint solver to obtain input seeds for its fuzzer. This unique amalgamation allows it to conduct an exhaustive search for vulnerabilities, thereby deepening the analysis.
    \item \textit{sFuzz} \cite{sFuzz}: sFuzz employs a randomized approach for smart contract analysis, utilizing a strategy based on American Fuzzy Lop (AFL). It introduces randomness into transaction sequences, thereby simulating real-world attack scenarios. The tool also adopts an adaptive strategy for selecting input seeds, which are then fed into the prominent fuzzer. This layer of stochasticity adds robustness to the vulnerability detection process.
    \item \textit{Slither} \cite{Feist2019}:  Slither employs IR through SlithIR and utilizes Single Static Assignment (SSA) for a more structured code analysis. Its structured approach facilitates streamlined code analysis, making it a highly valuable tool for secure contract development. Additionally, Slither aims to improve user understanding of contracts and assists with code reviews, broadening its utility.
    \item \textit{Conkas} \cite{Veloso2021}: Conkas offers a unique hybrid approach by combining IR and symbolic execution. It employs Rattle's IR \cite{rattle} technique to elevate bytecode to an intermediate representation. Instructions are converted to SSA form and transitioned from a stack form to a register form. A control-flow graph (CFG) is then constructed, over which symbolic execution iterates to generate traces. By identifying traces leading to vulnerabilities, it provides insights into the potential attack paths, enabling more targeted mitigation strategies.
    \item \textit{GNNSCVD} \cite{Zhuang2021}: GNNSCVD employs Graph Neural Networks (GNNs) to represent smart contract functions as graphs. During its training phase, these networks are fed a large number of normalized graphs from vulnerable contracts. The trained model is then used to absorb new graphs and assign vulnerability detection labels. 
    \item \textit{Eth2Vec} \cite{Ashizawa2021}: Eth2Vec utilizes machine learning techniques, specifically neural networks designed for natural language processing, to focus on code similarity analysis for detecting vulnerabilities. It compares target Ethereum Virtual Machine (EVM) bytecodes with known vulnerability patterns. Its learning-based approach allows for continuous adaptation and improvement.
    \item \textit{Solhint} \cite{solhint}: Solhint is a static analysis tool that checks code for potential issues and enforces best practices. It converts the contract source code into an Abstract Syntax Tree (AST), which enables Solhint to analyze the contract's logic and identify vulnerabilities, style violations, and potential optimizations by traversing the tree to assess each component of the code.
    \item \textit{SmartCheck} \cite{TikhomirovVITMA18}: SmartCheck transforms Solidity source code into an XML parse tree using ANTLR\cite{ANTLR} and a custom Solidity grammar as an IR. It then detects vulnerabilities by running XPath queries on this IR. The XML-based IR enables complete code coverage, with each element accessible via XPath.
    \item \textit{Maian} \cite{NikolicKSSH18}: Maian is a symbolic analysis tool that detects vulnerabilities by examining execution traces across multiple invocations. It categorizes contracts as "greedy," "prodigal," or "suicidal," depending on the type of flaw: locking funds indefinitely, leaking them to arbitrary users, or being killable by anyone.

  \end{itemize}

\subsection{Evaluation Criteria}
\label{criteria}
To systematically and objectively assess the quality of the selected analysis tools for smart contracts, we have developed a set of criteria grounded in the internationally recognized ISO/IEC 25010 standard \cite{Cai2020}. This standard provides a comprehensive framework for evaluating software product quality across eight distinct characteristics. Tailoring this framework to the specific needs and objectives of smart contract analysis, we focused our evaluation on the following four most pertinent characteristics:

\subsubsection{\textbf{Functional Suitablity}} 
Functional suitability concerns whether a software product meets functional requirements.
In our study, it can be measured by whether analysis tools meet functional desired requirements specified by users. For analysis tools, the most important functional requirement is the ability to find vulnerabilities in given smart contracts. 
Vulnerability detection can be seen as a binary classification problem, i.e., the tool has to predict whether or not a given contract contains a specific vulnerability. The task here is binary because there are only two possible outcomes: the presence or absence of the vulnerability. These outcomes fall into one of the following four categories:

\begin{itemize}
  \item True Positive (TP): Instances where the tool correctly identifies a contract as having a vulnerability when it does.
  \item False Positive (FP): Instances where the tool incorrectly identifies a contract as having a vulnerability when it does not.
  \item False Negative (FN): Instances where the tool fails to identify a contract as having a vulnerability when it actually does.
  \item True Negative (TN): Instances where the tool correctly identifies a contract as not having a vulnerability. 
\end{itemize}

The accuracy, which measures the proportion of correct predictions (both positive and negative) out of all predictions, can indeed be calculated with the formula you've provided:

\begin{equation}
  Accuacy = \frac{TP+TN}{TP+FP+FN+TN}
\end{equation}

This metric provides a straightforward measure of the tool's overall performance in correctly identifying both vulnerable and safe contracts.

F1 score is another metric in binary classification problems and is particularly useful when the data are imbalanced. It is the harmonic mean of precision and recall (sensitivity). 

\begin{itemize}
  \item Precision: The percentage of positive identifications which were actually correct. A model that produces no false positives has a precision of 1.0.
  \item Recall: The percentage of actual positives which were correctly identified. A model that produces no false negatives has a recall of 1.0.
\end{itemize}

The formula for the precision, recall and F1 score is:

\begin{gather}
	Precision = \frac{TP}{TP+FP} \\
	Recall = \frac{TP}{TP+FN} \\
    F_1= 2 * \frac{Precision*Recall}{Precision+Recall}
\end{gather}
where an F1 score is considered perfect when it is 1, while the model is a total failure when it is 0. 

Given that our datasets might exhibit class imbalance, the F1-score offers a balanced measure of the tool's precision and recall, providing insight into how well the tool performs on detecting vulnerabilities (the positive class). Therefore, when evaluating the tool's performance in terms of Functional Suitability, the average F1 score is used to quantify its capability. The F1 score here only considers the types of vulnerabilities that the tool is capable of detecting, avoiding overlap with the Usability(\ref{Usability}) metric.

\begin{gather}
    S_f = 2 * \frac{Precision_{avg} * Recall_{avg}}{Precision_{avg}+Recall_{avg}}
\end{gather}

\subsubsection{\textbf{Performance Efficiency}}
Performance efficiency concerns whether a software product can effectively make use of given resources. In the context of our study, this is quantified by examining the execution time required by various smart contract analysis tools.  

To gauge the performance efficiency of the selected analysis tools, we are interested in measuring the efficiency of various smart contract analysis tools by evaluating the average execution time for processing a contract. We first calculate the total execution time ($T_{total}$) and the count of validly analyzed contracts ($c_{vaild}$), excluding those that were terminated due to timeouts or Docker-related issues. The average execution time per contract ($T_{avg}$) is then determined using these values.

\begin{gather}
	T_{avg} = \frac{T_{total}}{c_{vaild}}
\end{gather}

To provide a standardized measure of performance efficiency, we scale the average time metric to a range from 0 to 1. Utilizing the min-max normalization method, we compute the performance time score, $S_{e}$, as given by:

\begin{gather}
	S_{e} = 1- \frac{T_{avg}-T_{min}}{T_{max}-T_{min}}
\end{gather}
where $T_{min}$ and $T_{max}$ denote the minimum and maximum values of $T_{avg}$ among the selected tools, respectively. In the context of our study, shorter average times are more favorable, prompting us to invert the normalized value. Thus, a lower $T_{avg}$ corresponds to a score closer to 1, while a higher $T_{avg}$ results in a score nearing 0.
This approach ensures that the normalized score accurately reflects the efficiency of the tools, with higher scores indicating more desirable performance.

\subsubsection{\textbf{Compatibility}}
Compatibility concerns whether a software product can consistently perform its functions while exchanging. 
In our study, we measure compatibility based on an analysis tool's ability to operate across multiple versions of the Solidity programming language. 
The Solidity language is constantly evolving, and new versions may introduce new features, syntax, or changes that may affect the behavior of smart contracts. Therefore, it is essential for analysis tools to stay abreast with the latest Solidity versions to sustain their effectiveness. Tools unable to analyze contracts composed in newer Solidity versions might be restricted in their vulnerability detection capabilities or could yield inaccurate results, potentially leading to false positives or negatives. Hence, compatibility with a wide range of Solidity versions is a crucial criterion in assessing a tool's utility and future-proofing potential. 

To quantify this compatibility, we introduce a metric to evaluate the compatibility of various analysis tools with different Solidity versions. Recognizing that the relevant Solidity versions range from 0.4.x to 0.8.x, we normalize the compatibility version value within this range to facilitate comparison. The compatibility score, $S_{u}$, is calculated as:

\begin{gather}
	S_{c} = \frac{S_{com} - S_{l}}{S_{h}-S_{l}}
\end{gather}
where $S_{com}$ represents the highest Solidity version the tool can analyze, while $S_{l}$ and $S_{h}$ donte the lowest and highest versions, respectively.

By scaling the compatibility in this manner, we provide a standardized measure that reflects the tool's ability to work with various Solidity versions, with a score of 0 indicating compatibility with only version 0.4.x, and a score of 1 indicating compatibility with the latest version 0.8.x. This metric offers insights into how well a tool can adapt to changes and updates in the Solidity language, which is crucial for its applicability in different development environments. In analysis tools, Solidity versions are backward compatible.

\subsubsection{\textbf{Usability}}
\label{Usability}
Usability concerns whether users can effectively and efficiently use a software product to complete tasks. In our study, we measure usability by observing whether analysis tools can cover more vulnerability types. 
To standardize this metric and ensure comparability across tools, we normalize the vulnerability type coverage score, scaling it to a range between 0 and 1. This normalized value, denoted as $S_{u}$, calculates the proportion of vulnerabilities detected by the tool relative to the total number of vulnerabilities examined in the experiment. The normalization is carried out using the following formula:

\begin{gather}
	S_{u} = \frac{S_{cov}}{S_{selected}}
\end{gather}
where $S_{cov}$ represents the raw vulnerability coverage score for the tool, and $S_{selected}$ indicates the total number of vulnerability types included in our evaluation. In our study, $S_{selected}$ is set to 10, corresponding to the ten vulnerability types we focus on.

By expressing the coverage in this way, we create a consistent and interpretable metric that facilitates the comparative evaluation of the selected tools' capabilities in identifying different vulnerability types.

\subsubsection{\textbf{Overall Assessment}}
In evaluating the analysis tools, it is essential to consider multiple dimensions of quality, including functional suitability, performance efficiency, compatibility, and usability. To create an overall score that balances these key aspects, we employ a weighted sum model, a widely recognized method for combining diverse evaluation metrics into a single, unified score.
We assign $\alpha$, $\beta$, and $\gamma$ as the weights corresponding to the importance of functional suitability, performance efficiency, and compatibility, respectively, in our specific context. The usability score is weighted with $1-\alpha-\beta-\gamma$. The formula for the overall score, $S_{over}$, is given by:

\begin{equation}
  S_{over} = \alpha \times S_{f} + \beta \times S_{e} + \gamma \times S_c + (1-\alpha-\beta-\gamma) \times S_u
\end{equation}

Thus, we have transformed the tool evaluation problem into a MCDM problem, with the key challenge being the determination of the weights for each criterion. To enhance the scientific and rational basis of the decision-making process and reduce bias, we have employed two methods—entropy weight method and analytic hierarchy process—to determine the weights.

\subsubsection{\textbf{Entropy Weight Method(EWM)}}
\label{EWM}
EWM is a technique commonly used in multi-criteria decision analysis to determine the weights of various evaluation criteria. This method is based on the principle of information entropy, which helps assess the importance of each criterion by measuring the amount of information it provides. The core idea is that the higher the entropy, the more dispersed the information provided by the criterion, and thus the lower its weight should be. Conversely, the lower the entropy, the more concentrated the information, leading to a higher weight for that criterion.

The advantages of EWM are primarily reflected in the following two aspects: \textcircled{1} Reduce Subjectivity. By calculating the entropy value of each criterion, the method reduces the influence of human subjectivity in the weight allocation, thereby increasing the objectivity of the decision-making process. \textcircled{2} Adaptability. The method can be applied to various decision-making problems, especially when there is a lack of prior knowledge or experience regarding the relationships between the evaluation criteria.

The specific process of EWM is as follows:
\paragraph{Data Standardization}
The first step is to standardize the raw data to eliminate the effect of different units or scales. In this study, we use the range normalization method for data standardization. The standardized data \(x'_{ij}\) for the \(i\)-th alternative and \(j\)-th criterion can be calculated as:

\begin{gather}
\label{11}
x'_{ij} = \frac{x_{ij} - \min(x_j)}{\max(x_j) - \min(x_j)}
\end{gather}

\paragraph{Calculate the Entropy Value of Each Criterion}

The entropy value for each criterion \(j\) is calculated to measure its dispersion or variability. The entropy \(E_j\) for criterion \(j\) is computed as:

\begin{gather}
\label{12}
E_j = -k \sum_{i=1}^{m} p_{ij} \ln(p_{ij})
\end{gather}

where \(p_{ij}\) is the proportion of the \(i\)-th alternative under the \(j\)-th criterion after standardization, and \(k\) is a constant, typically \(k=\frac{1}{\ln(m)}\), with \(m\) being the total number of alternatives.

\paragraph{Calculate the Weights for Each Criterion}
The weight \(w_j\) for each criterion is determined based on its entropy value. The weight is computed as:

\begin{gather}
\label{13}
w_j = \frac{1 - E_j}{\sum_{j=1}^{n} (1 - E_j)}
\end{gather}

where \(w_j\) is the weight for the \(j\)-th criterion, and \(n\) is the total number of criteria.

\subsubsection{\textbf{Analytic Hierarchy Process(AHP)}}
\label{AHP}
AHP is a popular multi-criteria decision-making method that breaks down complex decision problems into hierarchical levels. This method helps decision-makers make systematic comparisons between criteria at different levels and assign weights, leading to a final decision. AHP combines both qualitative and quantitative judgment to create a more scientific and systematic decision-making process.

The advantages of AHP are primarily reflected in the following two aspects: \textcircled{1} Simplification of Complex Problems. By decomposing the decision problem into hierarchical levels, AHP simplifies the analysis of each factor, making it easier for decision-makers to focus on individual elements. \textcircled{2} Integration of Qualitative and Quantitative Analysis. AHP can handle both qualitative judgments and quantitative data, providing a comprehensive approach to decision-making that considers both subjective and objective factors.

The specific process of AHP is as follows:

\paragraph{Constructing the Hierarchical Structure}
The first step is to decompose the decision problem into a hierarchical structure. Typically, the structure includes the goal level, criteria level, and alternative level.
\paragraph{Pairwise Comparison}
At each level, pairwise comparisons are made between the factors to assess their relative importance. These comparisons are generally made using a scale from 1 to 9, where 1 means equally important and 9 means one factor is extremely more important than the other.

\paragraph{Constructing the Pairwise Comparison Matrix}
The results of the pairwise comparisons are placed in a matrix. For two factors \(i\) and \(j\), if factor \(i\) is more important than factor \(j\), a value \(a_{ij}\) is assigned to the matrix. Conversely, if factor \(j\) is more important than factor \(i\), the reciprocal value \(a_{ji} = \frac{1}{a_{ij}}\) is assigned.
The judgment matrix \(A\) is formed as:

\begin{gather}
\label{AHP_matrix}
A = \begin{pmatrix}
1 & a_{12} & a_{13} & \cdots & a_{1n} \\
\frac{1}{a_{12}} & 1 & a_{23} & \cdots & a_{2n} \\
\frac{1}{a_{13}} & \frac{1}{a_{23}} & 1 & \cdots & a_{3n} \\
\vdots & \vdots & \vdots & \ddots & \vdots \\
\frac{1}{a_{1n}} & \frac{1}{a_{2n}} & \frac{1}{a_{3n}} & \cdots & 1
\end{pmatrix}
\end{gather}

\paragraph{Weight Calculation}

Once the judgment matrix is constructed, weights for each factor are calculated. The most common methods for calculating weights are the eigenvector method or the geometric mean method. The weight for the \(i\)-th factor, \(w_i\), is calculated by solving the following equation:
\begin{gather}
A \cdot w = \lambda_{\max} \cdot w
\end{gather}
where \(A\) is the judgment matrix, \(w\) is the weight vector, and \(\lambda_{\max}\) is the largest eigenvalue of the matrix.

\paragraph{Consistency Check}
AHP requires that the judgments made during the pairwise comparisons be consistent. To check the consistency, the consistency ratio \(CR\) is calculated as follows:
\begin{gather}
CR = \frac{CI}{RI}
\end{gather}
where \(CI\) is the consistency index, and \(RI\) is the random consistency index. If \(CR \leq 0.1\), the judgment matrix is considered consistent. If \(CR > 0.1\), the judgments should be revised.

\section{Benchmarking Dataset}
\label{evaluation}

One significant challenge in evaluating analysis tools is how to obtain a substantial number of amounts of vulnerable smart contracts for the evaluation. Despite the existence of numerous open-source analysis tools, comparing and reproducing these tools often proves difficult due to the scarcity of publicly accessible datasets. While several researchers have published their datasets, these have limitations such as unreasonable vulnerability taxonomy and small sample sizes of vulnerable contracts.

Recognizing the need for standardized benchmarks to facilitate effective evaluation, we have assembled a benchmark comprising two distinct datasets: a labelled dataset and a scaled dataset. These datasets are publicly accessible at our GitHub repository\footnote{https://github.com/bit-smartcontract-analysis/smartcontract-benchmark} and have been carefully constructed to address the limitations of existing datasets.

\subsection{Limitations of Existing Taxonomy:}

Currently, there are two popular taxonomies in the Ethereum community: Decentralized Application Security Project (DASP) TOP 10 and Smart Contract Weakness Classification (SWC). 

DASP Top 10 lists 10 types of vulnerabilities, which are used by a framework called SmartBugs \cite{Durieux2020} for analyzing Ethereum smart contracts. However, there are some issues: 
\textcircled{1}Version limitations. Since DASP was last updated in 2018, vulnerabilities such as \texttt{Short Addresses} have been addressed by the Solidity 0.5.0 and later versions of the compiler.
\textcircled{2}Inappropriate granularity. For vulnerabilities like access control and denial of service, the granularity of classification is too broad. Instances such as using \texttt{tx.origin} for caller authentication, causing ether leakage through the \texttt{delegatecall}, or the use of a protected self-destruct function leading to permanent ether freezing, all fall under access control vulnerabilities. As for denial of service vulnerabilities, they could be triggered by in-loop exceptions, unexpected interruptions, or gas fees exceeding limits. But in fact the detection reports of the tools tend to focus on finer granularity and cannot fully cover vulnerabilities like access control and denial of service.

The SWC includes 37 types of weaknesses. As mentioned in \cite{li2024static}, many SWC entries focus on contract code quality rather than vulnerabilities, such as SWC-135: Code With No Effects and SWC-102: Outdated Compiler Version. Therefore, this taxonomy is also not the best practice for evaluating smart contract vulnerability detection tools.

\subsection{Selection of Vulnerability Types:}
\label{Vulnerability Types}
A recent survey \cite{wei2024survey} classified blockchain smart contract vulnerabilities based on the Common Weakness Enumeration (CWE) rules, with granularity ranging from four primary root causes to 14 secondary causes, and then to 40 specific vulnerabilities. After filtering out the vulnerabilities relevant to Ethereum smart contracts, we selected 10 specific vulnerabilities as the types used for tool evaluation in this paper, based on  their frequency in the existing smart contract landscape \cite{Tolmach2022, Saad2020}.

In the following detailed descriptions, we delve into the nature and potential impact of each selected vulnerability (V1 to V10), shedding light on the underlying mechanisms that make them exploitable, and the consequences they may have on the smart contract functionality.
\begin{itemize}
  \item \textbf{Reentrancy (V1):} This vulnerability occurs when a contract calls an external contract, and the called contract then calls back into the calling contract before the first invocation is finished. This can lead to unexpected behaviour, particularly if the state of the contract was changed during the external call. 
  \item \textbf{Arithmetic (V2):} This occurs when an arithmetic operation generates a value that exceeds the range that can be represented within the fixed number of bits designated for integers in the EVM. This can lead to unexpected behaviour, as the EVM does not automatically check for these overflows or underflows.
  \item \textbf{Unchecked Send (V3):} This vulnerability happens when the call fails accidentally or an attacker forces the call to fail. It is also described as \textit{unhandled exceptions}, \textit{exception disorder}, or \textit{unchecked low-level call}.
  \item \textbf{Unsafe Delegatecall (V4):} This vulnerability rises from the \textit{DELEGATECALL} instruction, which allows a contract to dynamically load code from another contract at runtime. This can be exploited by an attacker to execute arbitrary code within the context of the calling contract, potentially leading to a loss of funds or other undesirable outcomes.
  \item \textbf{Transaction Ordering Dependence (V5):} This vulnerability, is also described as TOD, arises when a contract's behaviour depends on the order of transactions. Malicious actors, including miners, can potentially manipulate the order of transactions within a block to their advantage.
  \item \textbf{Time Manipulation (V6):} This vulnerability arises when smart contracts rely on the timestamp information from blocks.  However, because miners have some discretion over the timestamps of the blocks they mine, they could potentially manipulate these to their advantage, affecting the behaviour of contracts that depend on them. 
  \item \textbf{Bad Randomness (V7):} This vulnerability pertains to the flawed generation of random numbers within smart contracts. Random numbers often influence the decisions or outcomes of contract functionalities. If the process of random number generation is compromised, adversaries may predict the contract's outcome, leading to potential exploitation.
  \item \textbf{Authorization through tx.origin (V8):} This vulnerability arises when the \textit{tx.origin} variable is exploited by attackers. If contract authorization relies solely on \textit{tx.origin}, an attacker-controlled contract can trigger a transaction, tricking the victim contract into believing that the action was initiated by the original user address. This could result in unauthorized actions, such as changing ownership or draining funds. 
  \item \textbf{Unsafe Suicide (V9):} This vulnerability manifests when the \textit{SELFDESTRUCT} function is improperly secured and subsequently exploited by attackers. The \textit{SELFDESTRUCT} function in Ethereum allows a contract to be removed from the blockchain, effectively "killing" it. If this function isn't adequately protected, an attacker could potentially trigger it, destroying the contract prematurely. 
  \item \textbf{Gasless Send (V10):} This vulnerability occurs when there's an insufficient amount of gas to carry out an external call, resulting in the reversion of the transaction. This issue becomes particularly significant in scenarios involving multiple calls executed within a single transaction. An unsuccessful external call can disrupt the intended flow of the contract, potentially leading to unexpected consequences.
\end{itemize}

\subsection{Labelled Dataset:} 
The labelled dataset are created based on four primary factors:
\begin{itemize}
  \item {Code Size}: The dataset encompasses contracts of varying sizes to be indicative of the diversity in real-world smart contract domains. This includes small-scale contracts consisting of mere dozens of lines of code as well as more complex contracts that extend to hundreds of lines.
  \item {Functionality}: Our dataset incorporates contracts with a wide range of functionalities, such as tokens, wallets, and games, to capture the multifaceted nature of smart contract applications.
  \item {Annotated Vulnerability}:  Each vulnerable case within the dataset is labelled with at least one type of vulnerability from the ten categories. These annotations serve as a clear benchmark for the evaluation of analysis tools.
  \item {Correct Instances}: To provide a more comprehensive evaluation landscape, the dataset also includes contracts that are deemed safe and free from any known vulnerabilities. These instances are crucial for assessing the false positive rates of the analysis tools.
\end{itemize}

The data we selected comes from the following sources: examples from the DASP\footnote{https://dasp.co/} and SWC Registry\footnote{https://swcregistry.io/}, Smartbugs-curated \cite{Durieux2020}, Not-so-smart-contracts \cite{NosoSmart}, SolidiFi benchmark \cite{ghaleb2020effective}, Smart-Contract-Benchmark-Suites \cite{ren2021empirical}, and several contract security-related blogs. And based on the above requirements, our selection of vulnerabilities is informed by rigorous analysis and consultation. 

In Table \ref{tools}, we present the capability of the selected tools to identify particular vulnerabilities (V1 to V10). A checkmark (\checkmark) in the respective vulnerability column indicates the specific vulnerabilities that each tool can detect. For example, while Securify can identify vulnerabilities V1, V3 to V5, and V8, VeriSmart is solely capable of detecting V2.

\begin{table*}[ht]
\caption{The Overview of Selected Tools}
\centering
\label{tools}
\begin{tabular}{|l | l| l | p{0.4cm}<{\raggedright}| p{0.4cm}<{\raggedright} | p{0.4cm}<{\raggedright} | p{0.4cm}<{\raggedright}| p{0.4cm}<{\raggedright}| p{0.4cm}<{\raggedright}| p{0.4cm}<{\raggedright}|p{0.4cm}<{\raggedright}|p{0.4cm}<{\raggedright}|p{0.4cm}<{\raggedright}|}
\hline
Tool & Method & Version & V1 & V2 & V3 & V4 & V5 & V6 & V7 & V8 & V9 & V10 \\
\hline \hline
Securify & SA & v2.0 & \checkmark & & \checkmark & \checkmark & \checkmark & \checkmark & & \checkmark & & \\
\hline
VeriSmart & FV & 2021 & & \checkmark & & & & & & & & \\
\hline
Mythril & SE & 0.24.7 & \checkmark & \checkmark & \checkmark & \checkmark & & \checkmark & \checkmark & \checkmark & \checkmark & \\
\hline
Oyente & SE & latest & \checkmark & \checkmark & \checkmark & & \checkmark & & & & & \\
\hline
ConFuzzius & FZ, SE & v0.0.2 & \checkmark & \checkmark & \checkmark & \checkmark & \checkmark & \checkmark & \checkmark & & \checkmark & \\
\hline
sFuzz & FZ & latest & \checkmark & \checkmark & & \checkmark & & \checkmark & & & & \checkmark \\
\hline
Slither & IR, SA & 0.10.4 & \checkmark & & \checkmark & \checkmark & & \checkmark & & \checkmark & \checkmark & \\
\hline
Conkas & IR, SE & \#4e0f256 & \checkmark & \checkmark & \checkmark & & \checkmark & \checkmark & & & & \\
\hline
GNNSCVD & ML & latest & \checkmark & & & & & \checkmark & & & & \\
\hline
Eth2Vec & ML & latest & \checkmark & \checkmark & & & & \checkmark & & & & \checkmark \\
\hline
Solhint & SA & 5.0.3 & \checkmark & & \checkmark & \checkmark & & \checkmark & & \checkmark & \checkmark & \\ 
\hline
SmartCheck & SA & 4.13.1 & \checkmark & \checkmark & \checkmark & \checkmark & & \checkmark & & \checkmark & & \checkmark \\
\hline
Maian & SE & \#3965e30 & & & & & & & & & \checkmark & \\
\hline
\end{tabular}
\begin{tablenotes}
\footnotesize
\item[1] In the table, the \textbf{\textit{Method}} field represents the specific technical approaches employed by the vulnerability analysis tools, where \textbf{\textit{SA}} denotes static analysis, \textbf{\textit{SE}} denotes symbolic execution, \textbf{\textit{FV}} denotes formal verification, \textbf{\textit{FZ}} denotes fuzzing, \textbf{\textit{ML}} denotes machine learning, and \textbf{\textit{IR}} denotes intermediate representation.

\end{tablenotes}
\end{table*}

The labelled dataset is carefully designed to incorporate the ten selected vulnerabilities (V1 to V10), ensuring that it encompasses a representative range of scenarios that mirror real-world use cases and potential vulnerabilities in smart contracts. Listing \ref{lst:Token} offers an example of a vulnerable contract affected by Reentrancy vulnerability. More specifically, the contract first calculates the withdrawal amount and initiates an external call to the attacker's contract using {\small \texttt{msg.sender.call.value(amountToWithdraw)}}. However, the state variable {\small \texttt{userBalances[msg.sender]}} is only updated after the external call is completed. This external call transfers control to the attacker's contract, allowing the attacker to recursively call {\small \texttt{withdrawBalance}} before the original transaction is completed.

For ease of understanding, we have included annotations at the beginning of the contract code. These annotations specify the source from which the contract is derived, the author of the contract, and the lines where vulnerabilities are located.

\begin{lstlisting}[language=Solidity, caption={Example source code contract  with vulnerabilities}, label={lst:Token}]
/*
 * @source: https://consensys.github.io/smart-contract-best-practices/known_attacks/
 * @author: consensys
 * @vulnerable_at_lines: 17
 */

pragma solidity ^0.5.0;

contract Reentrancy_insecure {

    // INSECURE
    mapping (address => uint) private userBalances;

    function withdrawBalance() public {
        uint amountToWithdraw = userBalances[msg.sender];
        // <yes> <report> REENTRANCY
        (bool success, ) = msg.sender.call.value(amountToWithdraw)(""); // At this point, the caller's code is executed, and can call withdrawBalance again
        require(success);
        userBalances[msg.sender] = 0;
    }
}
\end{lstlisting}

By focusing on these specific vulnerabilities that are commonly observed in existing smart contract landscapes, our labelled dataset enables a targeted and relevant evaluation of the analysis tools. This alignment not only enhances the relevance and applicability of our findings but also allows for a nuanced understanding of how different tools perform in detecting these common and critical vulnerabilities. It ensures that the results reflect the practical challenges that developers and security analysts face in the field, thereby making the conclusions of our study directly actionable for enhancing smart contract security. 

Finally, our labelled dataset comprises 389 test cases, of which 372 are identified as vulnerable and 17 are designated as safe. These statistics are detailed in Table \ref{statistics}. 
For each type of vulnerability, we specify the number of affected contracts along with their total lines of code (LoC). To ensure an accurate LoC count, we have removed comments, special characters, and labeled symbolics from our calculations. 
This comprehensive account offers a fine-grained view of the distribution and extent of various vulnerabilities.

\begin{table}[ht]
  \caption{Statistics on Labelled Dataset}
  \centering
  \label{statistics}
  \begin{tabular}{ l | l | l }
  \hline
  \textbf{Type} & \textbf{Number} & \textbf{LoC} \\
  \hline
  Reentrancy & 81 & 19,261\\
  Arithmetic & 65 & 15,227\\
  Unchecked send & 52 &4,036\\
  Unsafe delegatecall & 12 &786 \\
  TOD & 60&25,207 \\
  Time manipulation & 60 & 14,450 \\
  Bad randomness & 10 & 1,223\\
  tx.origin & 10 & 998\\
  Unsafe suicidal & 11 &1,259\\
  Gasless send & 11 &1,330\\
  Safe contracts & 17 &1,556\\ \hline
  \textbf{Total} & 389 &85,333\\
  \hline 
  \end{tabular}
\end{table}

\subsection{Scaled Dataset} 

Furthermore, we have constructed a scaled dataset that offers a more extensive evaluation environment with the following requirements:

\begin{itemize}
  \item{Real-World Relevance}: The large dataset incorporates actual contracts deployed on the Ethereum blockchain. This ensures that our evaluations reflect real-world usage patterns, thereby providing a realistic and in-depth assessment of the performance of each analysis tool.
  \item{Scalability}: The size and scope of the large dataset enable us to simulate conditions that closely resemble the scale of real-world deployments. This comprehensive approach enhances our ability to evaluate the scalability and robustness of the various analysis tools under consideration.
\end{itemize}

The data acquisition for the scaled dataset follows a different strategy. The first step was to collect the contracts from the Ethereum blockchain. Most Solidity source code are available on Etherscan\footnote{https://etherscan.io/}, which offers API for users to obtain the contract source code associated with its address. We used Google BigQuery\footnote{https://cloud.google.com/blog/products/data-analytics/ethereum-bigquery-public-dataset-smart-contract-analytics} to collect the Ethereum contract addresses that have at least one transaction. And the following BigQuery request is used to select all the contract addresses and count the number of transactions that are associated with each contract\footnote{https://console.cloud.google.com/bigquery?project=ethereum-378701}:

\begin{lstlisting}[language=SQL,  caption={Google BigQuery request for solidity smart contracts and corresponding transactions}]
  SELECT contracts.address, COUNT(1) AS tx_count
  FROM bigquery-public-data.crypto_ethereum.contracts AS contracts
  JOIN bigquery-public-data.crypto_ethereum.transactions AS transactions
  ON (transactions.to_address = contracts.address)
  GROUP BY contracts.address
  ORDER BY tx_count DESC
\end{lstlisting}

Note that Etherscan does not store all the source code for every Ethereum contract, and many contracts with different addresses may have the same content, which implies further processing for the obtained files. Therefore, we adopted and improved upon the filtering method mentioned in \cite{Durieux2020}. We remove the duplicates in the case where the MD5 checksums of several source files are identical after removing all the spaces, line breaks and comments. Removing duplicates ensures that our dataset contains only unique and distinct smart contracts, thereby enhancing the robustness and validity of our analysis. It also maximizes computational efficiency by preventing redundant analysis of identical contracts. Additionally, we filtered out source files lacking the \textit{pragma solidity} statement to avoid compilation errors, further improving the quality of our dataset. It should be noted that our scaled dataset is not entirely representative of the full spectrum of real-world contracts.

Finally, \textbf{20,000} unique Solidity contracts from 2016-10-25 15:24:27 UTC to 2023-02-23 01:37:47 UTC are selected for the scaled dataset. Compare to the labelled dataset, this dataset is considerably larger, encompassing \textbf{8,542,517} lines of code, providing a robust foundation for further analysis and tool evaluation (We also removed comments, special characters, and labeled symbolics).

These two datasets collectively enable a balanced and extensive evaluation of the selected smart contract analysis tools. By offering a combination of controlled test cases and real-world examples, we provide a benchmark for rigorous testing that can accurately gauge the effectiveness, efficiency, and adaptability of each tool.

\subsection{Hardware configuration}
The performance of all tools was assessed on a cloud-based virtual machine provided by Aliyun, ensuring a consistent and high-performance environment for a fair comparison. The system was equipped with an Intel(R) Core(TM) i5-13400 CPU, featuring 8 cores operating at a clock speed of 4.4 GHz, and was supplemented with 32 GB of memory. The operating system was a 64-bit version of Ubuntu 22.04 LTS. This high-performance setup ensured that each tool was evaluated under optimal processing conditions, leveraging the power of Intel Xeon Platinum 8350C processors with 8 vCPUs and 30GB of RAM.

\begin{table*}[]
  \caption{Functional Suitability Comparison for Analysis Tools}
  \centering
  \label{analysis_tools}
  \resizebox{1\textwidth}{!}{
  \begin{tabular}{|l | l | p{0.8cm}<{\raggedright}| p{0.8cm}<{\raggedright} | p{0.8cm}<{\raggedright} | p{0.8cm}<{\raggedright}| p{0.8cm}<{\raggedright}| p{0.8cm}<{\raggedright}| p{0.8cm}<{\raggedright}|p{0.8cm}<{\raggedright}|p{0.8cm}<{\raggedright}|p{0.8cm}<{\raggedright}|p{0.8cm}<{\raggedright}|}
  \hline
  Tool & Criteria  & \rotatebox{90}{Reentrancy} & \rotatebox{90}{Arithmetic} & \rotatebox{90}{Unchecked send} & \rotatebox{90}{Unsafe delegatecall} & \rotatebox{90}{TOD} & \rotatebox{90}{Time manipulation} & \rotatebox{90}{Bad randomnes} & \rotatebox{90}{tx.origin} &\rotatebox{90}{Unsafe suicide}&\rotatebox{90}{Gasless send}&Average\\
  \hline \hline
  \multirow{4}*{Securify} & Accuarcy & 0.449  & - & 0.957 & - & 0.247 & -& -& - &-&-&0.523\\
  \cline{2-13}
  & Precision  & 1 &- &1 &- &1 &- &- &-  &-&-&1\\
  \cline{2-13}
  & Recall  & 0.333  &- &0.942 &- &0.033 &- &- &-  &-&-&0.404\\
  \cline{2-13}
  & F1-score  & 0.5 &- &0.970 & -&0.06 &- &- &- &-&-&0.576\\
  \hline 
  \multirow{4}*{VeriSmart} & Accuarcy & -  & \cellcolor{gray!25}\textbf{0.951} & - & - & - & -& -& -& -& -&0.951\\
  \cline{2-13}
  & Precision  & - & 0.984 & - &- &- &- &- &-  &-&-&0.984\\
  \cline{2-13}
  & Recall  & - & 0.954 &-&- &- &- &- &-  &-&-&0.954\\
  \cline{2-13}
  & F1-score  & - &\cellcolor{gray!25}\textbf{0.969} &- &- & -& - &- &- &- &-&0.969\\
  \hline 
  \multirow{4}*{Mythirl} & Accuarcy & 0.704  & 0.585 & 0.870 & 0.815 & - & 0.637 & 0.778 & 0.778 & 0.741 & - &0.704\\
  \cline{2-13}
  & Precision  & 1 & 1 & 1 & 1&- &1 &1 &1  &1& -&1\\
  \cline{2-13}
  & Recall  & 0.672 & 0.477 & 0.827 & 0.5 &- &0.533 &0.4 &0.4  &0.3&-&0.582\\
  \cline{2-13}
  & F1-score  & 0.782 & 0.646 &0.905 & 0.667 &- & 0.696 & 0.571 & 0.571 & 0.462 & - &0.826\\
  \hline 
  \multirow{4}*{Oyente} & Accuarcy & 0.459  & 0.427 & 0.797 & - & 0.247 & -&- &- &-&-&0.470\\
  \cline{2-13}
  & Precision  & 1 & 0.909 & 1 &- &1 &- &- &-  &-&-&0.977\\
  \cline{2-13}
  & Recall  & 0.347 & 0.308 & 0.731 & -&0.033 &- &- & - &-&-&0.341\\
  \cline{2-13}
  & F1-score  & 0.514 &0.460 & 0.844 & - & 0.065 &- & -& -&-&-&0.506\\
  \hline 
  \multirow{4}*{ConFuzzius} & Accuarcy & 0.827  & 0.854 & 0.913 & 0.621 & 0.234 & 0.792 & 0.455 & - & \cellcolor{gray!25}\textbf{0.814} & -&0.719\\
  \cline{2-13}
  & Precision  & 0.985 & 1 & 0.96 & 1&0.571 & 1& 1& - &1&-&0.917\\ \cline{2-13}
  & Recall  & 0.803 & 0.815 & 0.923 & 0.083 & 0.067& 0.733& 0.143 & - & 0.5& -&0.632\\
  \cline{2-13}
  & F1-score  & 0.884 &0.898 & 0.941 & 0.154 & 0.119 & 0.846 & 0.25 & - & \cellcolor{gray!25}\textbf{0.667} &- &0.749\\
  \hline 
  \multirow{4}*{sFuzz} & Accuarcy & 0.347  & 0.476 & 0.740 & 0.759 & - &0.247 & \cellcolor{gray!25}\textbf{0.852} &- &- & 0.679 &0.472\\
  \cline{2-13}
  & Precision  & 1 &1 &1 &1 &- & 1& 1&  - &-&0.75&0.991\\
  \cline{2-13}
  & Recall  & 0.210 & 0.338 & 0.654 & 0.417 &- &0.033 &0.6 &- &-& 0.273&0.306\\
  \cline{2-13}
  & F1-score  & 0.347 & 0.506 & 0.791 & 0.588 &- &0.064 & \cellcolor{gray!25}\textbf{0.75} &- &-& 0.4 & 0.467 \\
  \hline 
  \multirow{4}*{Slither} & Accuarcy & \cellcolor{gray!25} \textbf{0.980}  & - & 0.899 & 0.793 & - & 0.974 & - & 0.889 & \cellcolor{gray!25}\textbf{0.814} & - &0.938\\
  \cline{2-13}
  & Precision  & 0.988 &- &0.979 &1 &- &1 &- &1  &1&-&0.991\\
  \cline{2-13}
  & Recall  &  0.988 & -&0.885 &0.5 &- &0.967 &- &0.700  &0.500&-&0.896\\
  \cline{2-13}
  & F1-score  & \cellcolor{gray!25} \textbf{0.988} & - & 0.930 & 0.667 & -& 0.983 & - & 0.824 & \cellcolor{gray!25}\textbf{0.667} & - & 0.941 \\
  \hline 
  \multirow{4}*{Conkas} & Accuarcy & 0.949  & 0.939 & \cellcolor{gray!25}\textbf{0.986} & - & \cellcolor{gray!25}\textbf{0.935} & 0.948 & -&- &-&-& 0.950\\
  \cline{2-13}
  & Precision  & 0.975 & 0.984&1 &- &1 &1 &- &-  &-&-&0.990\\
  \cline{2-13}
  & Recall  & 0.963 & 0.938& 0.981 &- &0.917 &0.933 & -&-  &-&-&0.947\\
  \cline{2-13}
  & F1-score  & 0.969 & 0.960 & \cellcolor{gray!25}\textbf{0.990} &- &\cellcolor{gray!25}\textbf{0.956} &0.964 &- &- &-&-&0.968\\ \hline
  \multirow{4}*{GNNSCVD} & Accuarcy &  0.316 & - &-  &-  &-  &  & 0.273 &- &-&-& 0.298\\
  \cline{2-13}
  & Precision  &  1 & -&- &- & -&- & 1& - & -& -&1\\
  \cline{2-13}
  & Recall  &  0.173 &- &-  &- &- &- & 0.067 & - &- &- & 0.127\\
  \cline{2-13}
  & F1-score  & 0.295 & - & - &- & -&- & 0.125 & -&- &- & 0.226\\
  \hline 
  \multirow{4}*{Eth2Vec} & Accuarcy &  0.520 & 0.561 & - & - & - & 0.351 & - &- &-& 0.679& 0.494 \\
  \cline{2-13}
  & Precision  &  0.972 & 0.968 & -&- & -& 1& - & - & -& 1& 0.980\\
  \cline{2-13}
  & Recall  &  0.432 & 0.462 & - & -& -& 0.167 & - & - &- &0.182 & 0.355\\
  \cline{2-13}
  & F1-score  & 0.598 & 0.625  & - & -&- & 0.286 & - & -&- &0.308 & 0.521\\
  \hline
  \multirow{4}*{Solhint} & Accuarcy &  0.673 & - & 0.870 & \cellcolor{gray!25}\textbf{0.828} & - & \cellcolor{gray!25}\textbf{1} & - & \cellcolor{gray!25}\textbf{0.963} & 0.786 & - & 0.832 \\
  \cline{2-13}
  & Precision  &  0.980 & - & 0.939 & 0.769 & - & 1& - & 1 & 1 & - & 0.967\\
  \cline{2-13}
  & Recall  &  0.617 & - & 0.885 & 0.833 & - & 1 & - & 0.900 & 0.455 & - & 0.796\\
  \cline{2-13}
  & F1-score  & 0.758 & - & 0.911 & \cellcolor{gray!25}\textbf{0.800} & - & \cellcolor{gray!25}\textbf{1} & - & \cellcolor{gray!25}\textbf{0.947} & 0.625 & - & 0.873\\
  \hline
  \multirow{4}*{SmartCheck} & Accuarcy &  0.265 & 0.280 & 0.986 & 0.690 & - & 0.896 & - & 0.963 & - & \cellcolor{gray!25}\textbf{0.786} & 0.589 \\
  \cline{2-13}
  & Precision  &  0.909 & 0.800 & 0.981 & 0.800 & - & 1& - & 1 & - & 0.778 & 0.910\\
  \cline{2-13}
  & Recall  &  0.123 & 0.123 & 1 & 0.333 & - & 0.867 & - & 0.900 & - & 0.636 & 0.488\\
  \cline{2-13}
  & F1-score  & 0.217 & 0.213  & 0.990 & 0.471 & - & 0.929 & - & 0.947 & - & \cellcolor{gray!25}\textbf{0.700} & 0.635\\
  \hline
  \multirow{4}*{Maian} & Accuarcy &  - & - & - & - & - & - & - & - & 0.750 & - & 0.750 \\
  \cline{2-13}
  & Precision  &  - & - & - & - & - & - & - & - & 1 & - & 1\\
  \cline{2-13}
  & Recall  &  - & - & - & - & - & - & - & - & 0.363 & - & 0.363\\
  \cline{2-13}
  & F1-score  &  - & - & - & - & - & - & - & - & 0.533 & - & 0.533\\
  \hline  
  \end{tabular}
  }
  \begin{tablenotes}
    \footnotesize               
    \item[1] -: a tool cannot detect this vulnerability
  \end{tablenotes} 
\end{table*}

\section{Experimental Results}

In this section, we present the experimental results of the 13 selected tools against the evaluation criteria defined in Section \ref{criteria} on the labelled dataset. This is followed by the running results of the 4 best-performing tools on the scaled dataset. 

\begin{figure*}[tp]
  \centering
  \includegraphics[width=5.5in]{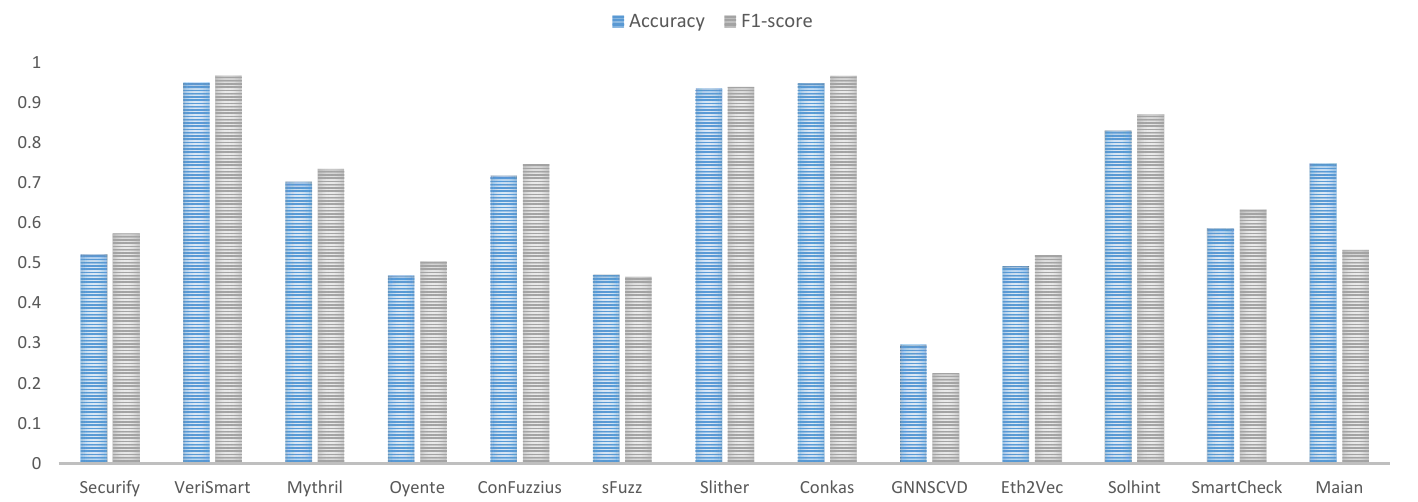}
  \caption{The Accuracy and F1-score of ten tools}
  \label{accuracy_F1}
\end{figure*}

\subsection{Functional Suitability}
To address RQ1, we first measure the functional suitability of the selected tools. We conduct a comprehensive performance evaluation of selected smart contract analysis tools, and the results are summarized in Table \ref{analysis_tools}. The table includes both well-established and emerging tools that leverage various analysis methods including formal verification, symbolic execution, fuzzing, and IR. 

We assessed each tool based on its accuracy, precision, recall, and F1-score across different vulnerability types (V1 to V10). Each cell in the table represents the corresponding evaluation metric for a specific tool against a specific vulnerability. The values provide the corresponding metric for each vulnerability type, with the last column showing the average across all vulnerabilities. Highlighted values (in shadow) indicate the highest scores for each vulnerability among the tools. 

The results illustrate a diverse range of performance among the different tools, demonstrating the strengths and weaknesses in each vulnerability. For example, Slither stood out in detecting Reentrancy and Unsafe Suicide vulnerabilities, with accuracy scores of 0.978 and 0.814, respectively. VeriSmart, on the other hand, was particularly superior in dealing with Arithmetic vulnerabilities, achieving an accuracy of 0.951, indicating its strength in ensuring arithmetic safety and mathematical consistency. 

Besides, we extracted the average accuracy and F1-score for each tool from the last column in Table \ref{analysis_tools} to further analyze the correlation between these two metrics. And the data is visually represented in Figure \ref{accuracy_F1}, which reveals a noticeable similarity in the trends exhibited by both accuracy and F1-score metrics across the various tools. 

Except for Maian, the absolute difference between the average Accuracy and F1-score of other tools does not exceed 0.075. Considering that Maian can only detect one type of vulnerability and that the distribution of vulnerability categories covered by the 13 tools is imbalanced, we ultimately decided to use the F1-score as the functional metric for evaluating vulnerability analysis tools.

\setlength{\tabcolsep}{4pt}
\begin{table*}[h]
  \begin{center}
  \caption{Performance for Automatic Analysis Tools}
  \label{analysis_performance}
  \begin{tabular}{|l | l|  l|  l |  l | l | l | l| l| l | l | l | l | l | }
  \hline
   & Securify & VeriSmart & Mythirl & Oyente & ConFuzzius & sFuzz & Slither & Conkas & GNNSCVD & Eth2Vec & Solhint & SmartCheck & Maian\\
  \hline
  $T_{total}$ & 1.3h & 1.4h  & 55.6h & 12min & 86.0h & 33.2h & \cellcolor{gray!25}5.3min &7.6h & 20.6min & 7.9min & 9.6min & 17.7min & 6.3h \\  \hline
  $T_{valid}$  & 350 & 82  & 329 & 167 & 349 & 134 &\cellcolor{gray!25}316 & 349 & 50 & 216 & 389 & 389 & 371 \\  \hline
  $T_{avg}$  & 14s & 63s & 609s & 5s & 887s & 289s & \cellcolor{gray!25}1.2s &78s& 24.76 & 2.2s & 1.5s & 2.7s & 61s  \\  \hline
  $S_{e}$  & 0.985  &0.931  & 0.318 & 0.996 & 0.007  & 0 & \cellcolor{gray!25}{1} & 0.913& 0.973 & 0.998 & 0.999 & 0.998 & 0.933\\  \hline

  \end{tabular}
  \end{center}
\end{table*}

\begin{mdframed}
  \textbf{Answer to RQ1: Does the analysis tools correctly analyze the smart contracts, providing accurate and reliable results?} 
  
  The results reveal that the target tools are capable of covering a wide spectrum of known vulnerabilities. For every vulnerability type, except Gasless send, there exists at least one tool that achieves an accuracy rate above 80\%. Among them, there are three tools capable of detecting six or more types of vulnerabilities with an average F1-score exceeding 80\%: Mythril, Slither, and Solhint. Additionally, these three tools are the only ones among the 13 that have actively maintained code repositories.
\end{mdframed}

\subsection{Performance Efficiency}
To address RQ2, we evaluated the computational efficiency of the selected analysis tools, including the total execution time ($T_{total}$), the number of valid execution contracts ($C_{valid}$), the average execution time ($T_{avg}$), and the performance time score ($S_{e}$), as detailed in Table \ref{analysis_performance}. 

In conjunction with Table \ref{tools}, deeper insights can be uncovered. Tools utilizing static analysis techniques exhibit an average execution time of under 15 seconds, occupying four of the top six spots in terms of time efficiency. On the other hand, the fuzzing tools ConFuzzius and sFuzz demonstrate significantly lower execution efficiency, highlighting the inherent drawbacks of dynamic analysis. For tools employing symbolic execution techniques, the average execution time ranges from 5 seconds (Oyente) to 609 seconds (Mythril). This discrepancy arises from the internal implementation logic of these tools: the most critical challenge in symbolic execution is the state explosion problem. To address this, these tools implement different time and coverage limits, terminating execution as soon as the predefined criteria are met.

\begin{mdframed}
  \textbf{Answer to RQ2: How is the performance of the target analysis tools in terms of detection time?} 
  
  Our analysis uncovers a wide variation in average execution time across the target tools, ranging from 1.2s to 887s. This execution time is closely tied to the methodology employed by each tool. Tools utilizing fuzz testing tend to require more time, while those employing IR are typically faster. The choice between these methodologies can significantly impact the efficiency of the tool, depending on the specific context and needs of the analysis task.
\end{mdframed}

\subsection{Compatibility and Usability}
To address RQ3, we assessed both the compatibility and usability of the selected smart contract analysis tools.  Our results are summarized in Table \ref{analysis_cap}, where we present four key indicators: compatibility version ($S_{com}$), compatibility version score ($S_{c}$), vulnerability coverage ($S_{cov}$), converage score ($S_u$).

We observe a variability in compatibility across the tools. Tools such as Mythril, ConFuzzius and Solhint exhibit excellent compatibility with the latest 0.8.x version of the Solidity programming language, thus receiving a full compatibility score ($S_{c}=1$). This is indicative of their up-to-date maintenance and adaptability to the evolving smart contract landscape. In contrast, tools like Oyente and sFuzz show limitations in their compatibility, supporting only the older 0.4.24 version of Solidity. This is also reasonable, as the last update for the former was in 2020, and for the latter in 2019, while Solidity 0.6.0 version was not released until 2020.

In the context of vulnerability coverage, Mythril and ConFuzzius stand out for their extensive capabilities. Both tools can detect up to 8 different types of vulnerabilities, garnering them a high coverage score ($S_{u}=0.8$). This breadth of vulnerability coverage implies a more comprehensive security audit, making these tools particularly valuable for a wide-ranging evaluation of smart contracts.
On the other hand, some tools like VeriSmart demonstrate a more limited scope, achieving a lower coverage score. 

\begin{mdframed}
    \textbf{Answer to RQ3: To what extent do target analysis tools support different Solidity compiler version and vulnerability?} 
    
    Our evaluation indicates a diverse landscape in terms of compatibility and vulnerability coverage among smart contract analysis tools. Tools that maintain up-to-date compatibility with recent Solidity versions, like Mythril and ConFuzzius, offer broader applicability for current smart contract security assessments. On the other hand, tools like Oyente and sFuzz, which support older versions, hamper their adaptability to the evolving smart contract landscape.
    The vulnerability coverage scores suggests that while some tools offer a comprehensive analysis across multiple types of vulnerabilities, others may be more specialized but excel in their focus areas. Therefore, the choice of an analysis tool should be guided not only by its compatibility and coverage but also by the specific analytical requirements of the task at hand.
\end{mdframed}

\setlength{\tabcolsep}{4pt}
\begin{table*}[h]
  \begin{center}
  \caption{Compatibility and Coverage for Automatic Analysis Tools}
  \label{analysis_cap}

  \begin{tabular}{|l | l|  l|  l |  l | l |l | l| l| l| l| l| l| l| }
  \hline
  & Securify & VeriSmart & Mythirl & Oyente & ConFuzzius & sFuzz & Slither & Conkas & GNNSCVD & Eth2Vec & Solhint & SmartCheck & Maian\\
  \hline
  $S_{com}$ & 0.5.x & 0.5.x & 0.8.x & 0.4.19 & 0.8.x & 0.4.24 &0.8.x & 0.5.x & 0.8.x & 0.8.x & 0.8.x & 0.5.x & 0.8.x\\  \hline
  $S_{c}$  & 0.25 & 0.25 & 1 & 0 & 1 & 0 & 1 & 0.25 & 1 & 1 & 1 & 0.25 & 1\\  \hline
  $S_{cov}$ &  3 & 1 & 8 & 4 & 8 & 7 & 6 & 5 & 2 & 4 & 6 & 7 & 1\\  \hline
  $S_{u}$ &  0.3 & 0.1 & 0.8 & 0.4 & 0.8 & 0.7 & 0.6 & 0.5 & 0.2 & 0.4 & 0.6 & 0.7 & 0.1\\  \hline
  \end{tabular}
  \end{center}
\end{table*}

\subsection{Overall Effectiveness Evaluation}
To address RQ4, we provide a comprehensive assessment of the overall performance of the selected smart contract analysis tools. This evaluation is carried out by assigning different weightings to key dimensions functional suitability ($\alpha$), performance efficiency ($\beta$), compatibility ($\gamma$) and usability ($1-\alpha-\beta-\gamma$).

As mentioned in Section \ref{EWM} and \ref{AHP}, we adopted two methods for weight assignment(Table \ref{3weight}). For EWM, no prior knowledge is required, and a set of weights can be derived solely by following Equations \ref{11}-\ref{13}. It is evident that this method assigns weights of less than 0.2 to the indicators Functional Suitability and Performance Efficiency, while assigning the highest weight (0.331) to the Usability indicator. However, our perspective is that, when faced with the selection of various smart contract analysis tools, the vulnerability coverage of a single tool, represented by Usability, should be the least important factor. Instead, Functional Suitability should be the primary criterion, followed by Compatibility and Performance Efficiency.

\begin{gather}
A = \begin{pmatrix}
1 & 4 & 2 & 5\\
\frac{1}{4} & 1 & \frac{1}{2} & 3\\
\frac{1}{2} & 2 & 1 & 3\\
\frac{1}{5} & \frac{1}{3} & \frac{1}{3} & 1
\end{pmatrix}
\end{gather}

\begin{gather}
A = \begin{pmatrix}
1 & 2 & 2 & 4\\
\frac{1}{2} & 1 & 1 & 2\\
\frac{1}{2} & 1 & 1 & 3\\
\frac{1}{4} & \frac{1}{2} & \frac{1}{3} & 1
\end{pmatrix}
\end{gather}

Therefore, in the AHP method, we constructed two Pairwise Comparison Matrices based on Equation 14. Both matrices reflect the same relative importance of the indicators, but the first matrix generated larger weight differences (Table \ref{3weight}, row "AHP1").

\begin{figure*}[ht]
  \centering
  \includegraphics[width=5.5in]{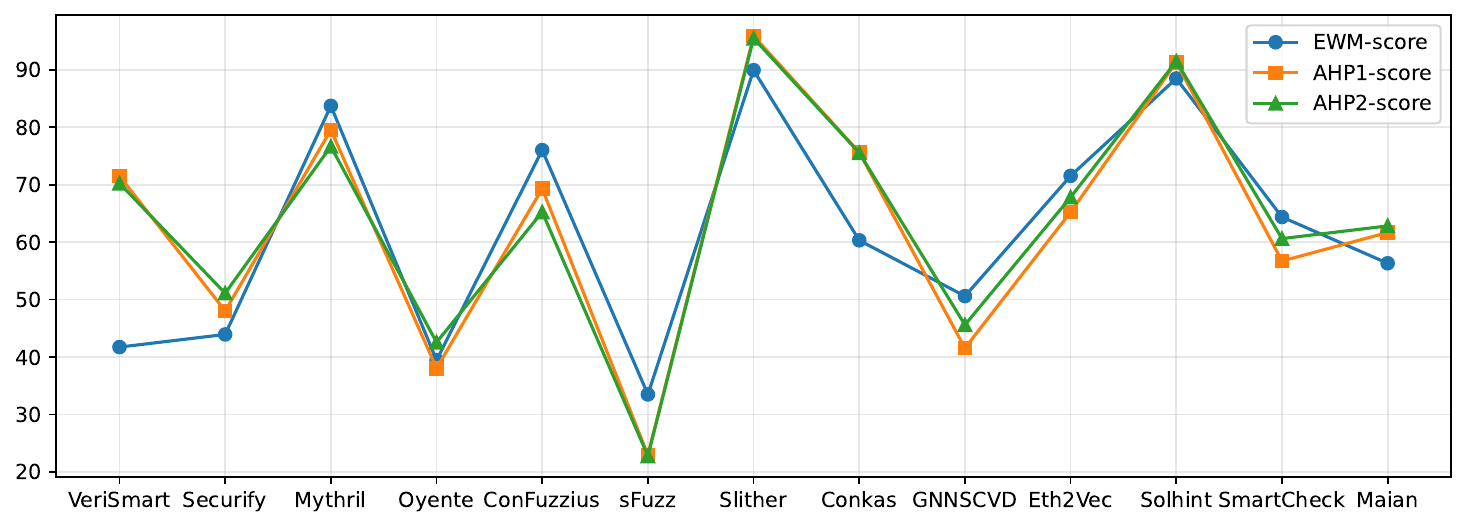}
  \caption{Overall effectiveness evaluation for different tools}
  \label{fig_overall}
\end{figure*}

Table \ref{overall} details the final scores of each tool under three different weight configurations and uses a line chart to visually enhance the comparison(Figure \ref{fig_overall}). Regardless of the decision-making method used for weight assignment, we found that Slither, Solhint, and Mythril consistently ranked among the top three tools in terms of scores. This validates the accuracy of our decision-making: these three tools are the only ones among the 13 smart contract analysis tools that are still actively updated and iterated by their communities. Furthermore, their GitHub star rankings are 1st, 4th, and 2nd, respectively, among the 13 tools (considering that Oyente was the first formal automated smart contract analysis tool, its 3rd place in GitHub stars is also reasonable).

For Verismart and Conkas, the significant discrepancies between their EWM-scores and the two AHP-scores can be explained as follows: Verismart is limited to detecting only arithmetic vulnerabilities, while Conkas only supports the analysis of Solidity contracts with version 0.5.x. However, EWM assigns relatively high weights to the Compatibility and Usability indicators, which disproportionately benefits these tools.

\begin{mdframed}
\textbf{Answer to RQ4: How to evaluate tools comprehensively, objectively, and effectively based on experimental data?}

We used two decision-making methods to assign weights to four indicators. The EWM method is entirely objective, while for AHP, we set up two Pairwise Comparison Matrices, resulting in three different sets of weights. These weights were then used to calculate the final scores for each tool. For most tools, the differences among the three scores were minimal. However, Verismart and Conkas showed significantly lower EWM-score compared to the two AHP-scores due to their weaknesses in specific indicators.

The tools that ranked in the top three in terms of scores also received broad recognition from the Ethereum community, further enhancing the credibility of our scoring framework. This method is comprehensive, objective, and efficient, providing a new approach for evaluating smart contract analysis tools in the future.

\end{mdframed}

\begin{table}[h]
  \begin{center}
  \caption{Weights of Each Indicator}
  \label{3weight}
  \begin{tabular}{| l | c | c | c | c |}
  \hline
  & \makecell{Functional\\Suitablity} & \makecell{Performance\\Efficiency} & Compatibility & Usability\\ \hline
  & $\alpha$ & $\beta$ & $\gamma$ & $1-\alpha-\beta-\gamma$\\ \hline  
  
  EWM  & 0.157 & 0.194 & 0.317 & 0.331\\  \hline
  AHP1 & 0.502 & 0.159 & 0.261 & 0.078\\  \hline
  AHP2 & 0.438 & 0.219 & 0.243 & 0.100\\  \hline
  \end{tabular}
  \end{center}
\end{table}

\begin{table*}[h]
  \begin{center}
  \caption{Overall Performance for Analysis Tools}
  \label{overall}
  \begin{tabular}{|l | l|  l|  l |  l | l |l | l| l| l| l| l| l| l| }
  \hline
  & Securify & VeriSmart & Mythirl & Oyente & ConFuzzius & sFuzz & Slither & Conkas &GNNSCVD &Eth2Vec & Solhint & SmartCheck & Maian\\
  \hline
  EWM  & 41.7 & 43.9 & 83.7 & 39.5 & 76.0 & 33.5 & \cellcolor{lightgray}{89.9} & 60.3 & 50.6 & 71.6 & 88.5 & 64.4 & 56.3  \\  \hline
  AHP1  & 71.5 & 48.0 & 79.5 & 38.1 & 69.3 & 23.0  & \cellcolor{lightgray}{95.9} & 75.7 & 41.6 & 65.2 & 91.3 & 56.7 & 61.7\\  \hline
  AHP2 & 70.3 & 51.1 & 76.7 & 42.6 & 65.3 & 22.8 & \cellcolor{lightgray}{95.5} & 75.5 & 45.6 & 67.8 & 91.5 & 60.6 & 62.8\\  \hline
  \end{tabular}
  \end{center}
    \begin{tablenotes}
    \footnotesize               
    \item[1] The final scores are retained to three decimal places and then multiplied by 100 for clarity.
  \end{tablenotes}
\end{table*}

\subsection{Real-world Evaluation} 
To answer RQ5, we focused on analyzing the scaled dataset of 20,000 smart contracts using the most overall effective tools based on our findings. Considering the diversity of analysis techniques and the scores of the tools, we selected tools Mythril, Slither, Conkas and ConFuzzius. Our primary goal in this part of the study was to identify the distribution of nine distinct types of vulnerabilities(excluding Vulnerability Gasless Send) in these contracts.

\begin{figure*}[ht]
  \centering
  \includegraphics[width=5.5in]{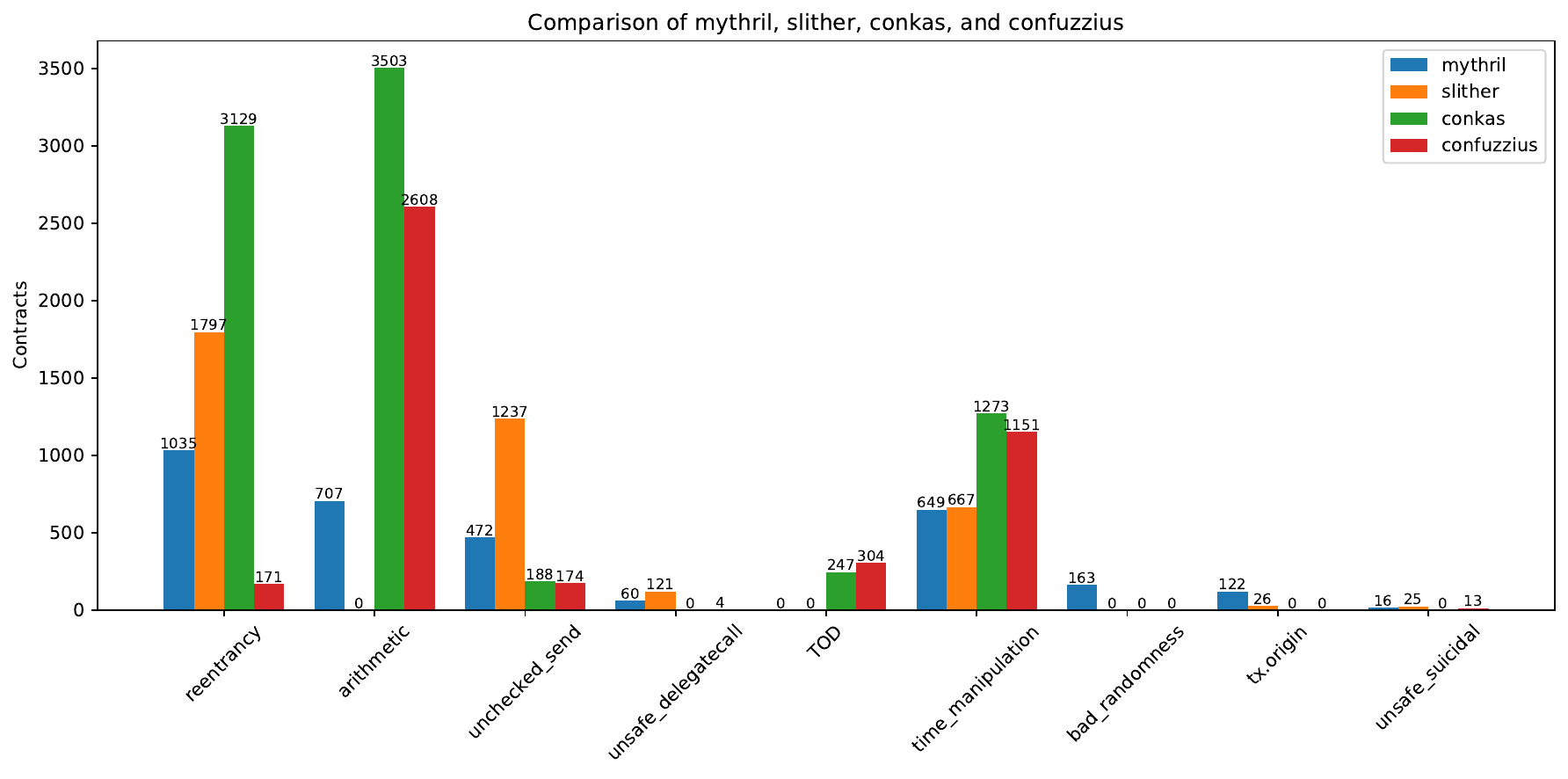}
  \caption{Real-world effectiveness evaluation}
  \label{fig_real_vulner}
\end{figure*}

As depicted in Figure \ref{fig_real_vulner}, the analysis uncovers a multifaceted landscape of security challenges within the contracts, where a value of 0 indicates that the tool has no ability to detect that particular vulnerability. Arithmetic vulnerability emerged as the most pervasive, with 707 instances identified by Mythril, 2608 by ConFuzzius, and 3503 by Conkas, underscoring a prevalent issue in managing numerical computations. The discrepancy between these numbers highlights a key difference in F1-score between the two tools (Mythril at 0.646, and Conkas at 0.960). In contrast, tx.origin and Unsafe suicidal vulnerabilities were minimally detected, numbering only 26 and 25 instances, respectively, thereby indicating their infrequent occurrence in different types of contracts. The notable presence of Reentrancy and Time manipulation vulnerabilities emphasizes the complexity of managing state and timing in smart contracts. Interestingly, although both Slither and Conkas performed best at detecting Reentrancy and Time manipulation vulnerabilities in our labelled dataset, their performance varied significantly in the scaled dataset. This discrepancy might stem from the limited variety of reentrancy contracts in the labelled dataset. Other vulnerabilities, such as unchecked send, unsafe delegatecall, and bad randomness, were found in moderate numbers, reflecting a wide array of potential security risks.

Moreover, we extended our investigation to analyze the evolution of specific vulnerabilities in the scaled dataset. For this analysis, we chose to combine the detection results from Mythril, Slither, and Conkas, as these tools were representative of the vulnerabilities detected.  
Initial observations reveal a marked increase in the total number of unique contracts towards the end of 2017, coinciding with Ether's peak value. Upon a detailed examination, two distinct trend groups become apparent. The first group includes Reentrancy, Arithmetic, Unchecked send, and Time manipulation vulnerabilities. This group exhibited a rapid growth in vulnerable contracts from October 2017 to October 2020, mirroring the surge in popularity and complexity of smart contracts during this period. The subsequent slowdown post-2021 may be attributed to several factors. First, the introduction of multiple security methods and newer Solidity versions effectively addressed previous security concerns. Second, the scope of application of Ethereum smart contracts appears to have become more standardized or fixed after 2021, potentially reflecting a maturation in the development practices and use cases for smart contracts. The second group, consisting of TOD, Unsafe delegatecall, Bad randomness, tx.origin, and Unsafe Suicidal vulnerabilities, remains relatively infrequent and seems to be readily controllable through various design precautions. These trends provide nuanced insights into the dynamic nature of smart contract vulnerabilities and reinforce the importance of tailored security measures to meet the evolving challenges.

\begin{figure}[tp]
  \centering
  \includegraphics[width=3.5in]{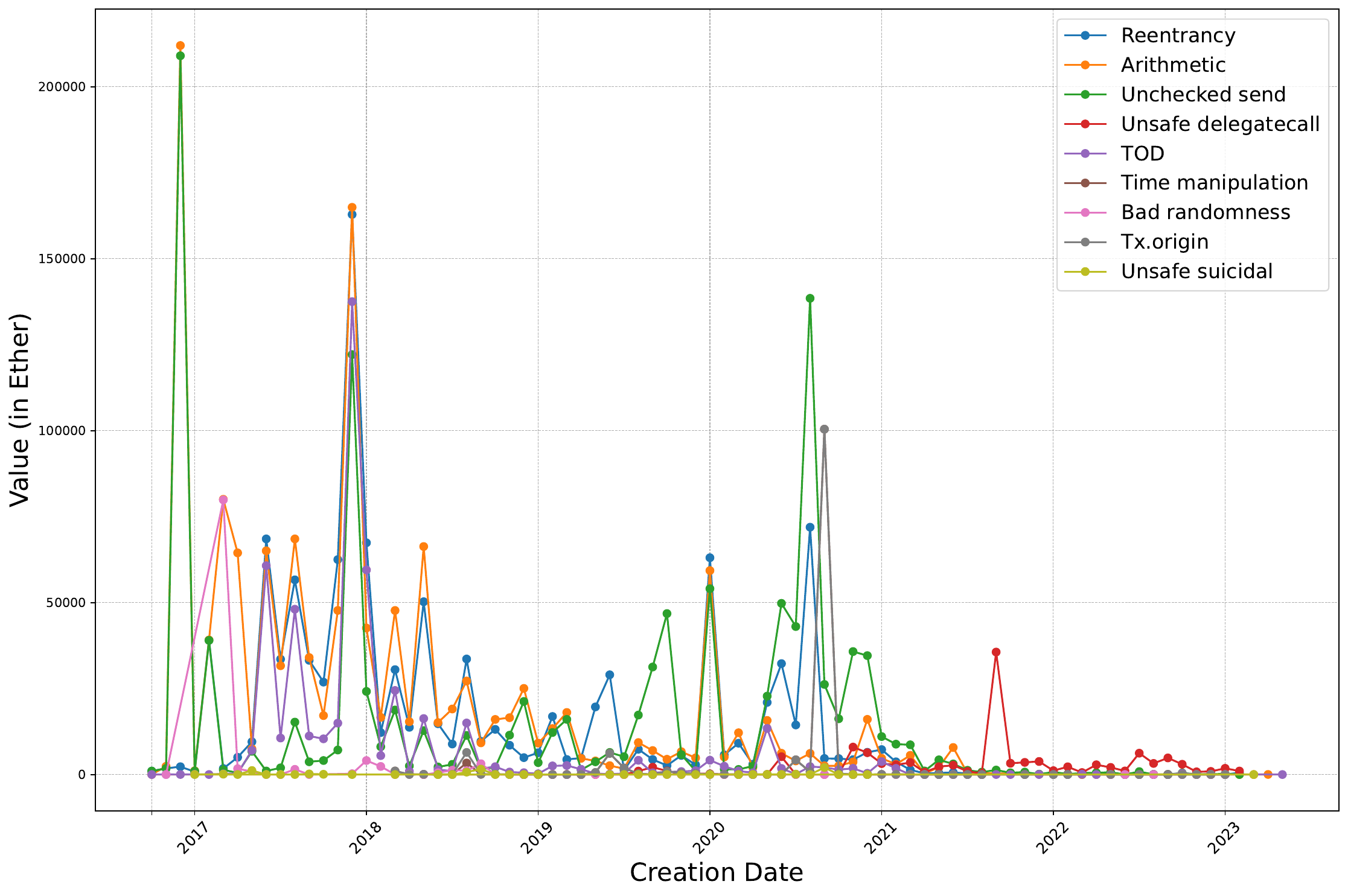}
  \caption{Value of vulnerable contracts over time}
  \label{fig_evolution_value}
\end{figure}

Additionally, we broadened our analysis to assess the financial impact of these vulnerabilities by investigating the transaction values linked to the vulnerable contracts. Specifically, this value is computed based on the transactions in which these vulnerable contracts were utilized. We sourced this transaction data directly from the Ethereum blockchain.  Figure \ref{fig_evolution_value} highlights the time-based variations in the financial value of contracts tied to different types of vulnerabilities. A general trend shows a decline in the value associated with vulnerable contracts over time. Upon detailed examination, certain vulnerabilities manifest heightened activity during specific timeframes. For example, contracts with Unsafe suicidal vulnerability once involved large sums (over 200,000 ether) but have dwindled to near zero. Arithmetic and TOD vulnerabilities were predominantly active prior to 2019. Unchecked send, which often pertain to large ether transactions, were prevalent until 2021. Unsafe delegatecall saw no significant ether-related issues until 2020 but experienced a sharp increase in 2021, due to increased use of the \textit{delegatecall} function for ether transactions. These patterns offer insights into evolving Solidity usage habits and associated financial risks.

\begin{mdframed}
  \textbf{Answer to RQ5: What is the outcome of tools on real-world contracts?} This analysis uncovers the complex landscape of smart contract security, identifying prevalent vulnerabilities and their progression over time. Upon evaluating the diversity of vulnerabilities identified by the specific tools, we find that Arithmetic and Reentrancy vulnerabilities are disproportionately prevalent in the Ethereum ecosystem. Moreover, we observe that unique contracts are being deployed at an accelerating rate, indicating a broader application of smart contracts. In contrast, the rate of new vulnerable contracts appearing has decelerated, revealing a growing awareness and increased emphasis on smart contract security.  Additionally, our assessment of the financial stakes involved in vulnerable contracts reveals periodic variations in security risks. These trends indicate that the gravity of different security issues shifts over time.
  These insights can assist developers and security professionals in strategically prioritizing and customizing their remediation efforts, ensuring that they effectively address the most common and pressing risks within the smart contract ecosystem.
\end{mdframed}

\section{Discussion}
\label{discussion}
\subsection{Summary of Findings}
Based on the above evaluation results, we have derived the following key findings:

\begin{itemize}
  \item{\textit{Recommondations:}} The evaluation of vulnerability analysis tools is a multidimensional task. We adopted the ISO/IEC 25010 standard to define the evaluation indicators, applied Multi-Criteria Decision Making  algorithms to assign weights to these indicators, and ultimately calculated the tool scores. This evaluation framework has been empirically validated as effective, and we identified some exceptionally outstanding tools that primarily employ symbolic execution and intermediate representation techniques. However, tools based on fuzzing techniques still have significant room for improvement, presenting an opportunity for researchers to address these gaps.
  \item{\textit{Inconsistency:}} The selection of an optimal tool heavily depends on the accuracy, execution time, specific types of vulnerabilities, and Solidity version. The effectiveness of a tool is not uniform across all vulnerability types. Therefore, being aware of each tool's strengths and weaknesses is critical for different scenarios. Tailoring the choice of tools to the nature of the project and the expected vulnerabilities can significantly enhance the effectiveness of the detection process.  
  \item{\textit{Combined tools:}} Utilizing multiple tools can enhance code quality and vulnerability detection. Leveraging the unique strengths of different tools, such as Mythril for known vulnerabilities and Slither for deep code analysis, offers a more comprehensive security assessment. This synergistic approach allows for a more nuanced analysis, potentially uncovering vulnerabilities that might be missed by a single tool. It promotes a more robust and secure development process, aligning with best practices in software security.
  \item {\textit{Trend of vulnerabilities:}} The analysis of real-world data shows a significant slowdown in vulnerability emergence after a rapid growth phase (October 2017 to October 2020). This trend likely stems from the increased economic value of smart contracts, the proliferation of security tools, and the standardization of Ethereum smart contract practices after 2021. The trend signifies a maturation in the smart contract ecosystem and an increased focus on security.
\end{itemize}

\subsection{Threats to Validity}

\begin{itemize}
\item \textit{Categorization of smart contract vulnerabilities:}
Different researchers may interpret and assess the severity and classification of vulnerabilities in varying manners. This subjectivity could introduce bias and potentially impact the validity of the comparative analysis in our evaluation. In an effort to counteract this threat, we have employed a systematic approach rooted in industry standards and best practices, as discussed in Section \ref{methodology}. Each vulnerability category was thoroughly scrutinized and discussed in our previous publication to ensure a uniform understanding and categorization.

\item \textit{Generality of evaluation datasets:} The applicability of our conclusions may be constrained by how accurately the datasets reflect real-world scenarios and diverse smart contract usage patterns. Inadequate coverage of various applications could limit the broader relevance of our findings. To address this, we collected a comprehensive and varied set of contract tests from diverse sources, striving to represent the real-world usage and diversity of smart contracts as faithfully as possible.

\item \textit{Focus on open-sourced tools:}  Our emphasis on open-sourced tools, rather than commercial products, may lead to selection bias, restricting the generalizability of the results. While open-source tools offer transparency and are commonly utilized in research, they might not represent the full spectrum of methodologies and performance characteristics found in commercial products.
\end{itemize}

\section{Conclusion}
\label{conclusion}
In this survey, we conducted an exhaustive evaluation of automated smart contract analysis tools, revealing significant insights into their efficacy, utility, and performance under different scenarios. We found that no single tool excels universally, i.e., each has its own set of advantages and drawbacks depending on the vulnerability type and evaluation metrics. Using the ISO/IEC 25010 standard as our assessment framework, we were able to offer a robust and systematic comparison of these tools, providing clear insights into their quality and performance efficiency. Our benchmark dataset serves as a valuable reference for both future researchers and industry practitioners, facilitating the evaluation of emerging tools. Our study also presents important observations on the evolution of security risks in real-world contracts.

Despite these insights, the smart contract landscape is evolving rapidly, introducing new functionalities and protocols that bring forth novel security challenges.
Future research should expand to encompass the investigation of more sophisticated or emerging vulnerabilities. This is particularly pertinent given the ongoing innovation and updates in smart contract development, which inevitably give rise to new and unforeseen security concerns. Moreover, the scope of smart contract research should broaden to include other programming languages. For instance, the growing interest in using programming languages such as Go and Rust, in addition to Solidity, for developing smart contracts calls for comprehensive analysis across these platforms.


\begin{table*}[htp]
  \begin{center}
  \caption{Vulnerability menu}
  \label{vulner_list_first}
  \begin{tabular}{|l |l | l| l|  }
  \hline
   \textbf{Cause} & \textbf{Vulnerability Type} & \textbf{Blockchain Platform} & \textbf{ID}\\ \hline\hline
   \multirow{2}*{Syntax Error} & Typographical Error & Ethereum & VE1 \\
  \cline{2-4}
  & Right-To-Left-Override control character & Ethereum & VE2 \\
  \hline
  \multirow{5}*{Version Issue} & Outdated Compiler Version & Ethereum & VE3 \\
  \cline{2-4}
  & Use of Deprecated Solidity Functions & Ethereum & VE4 \\
  \cline{2-4}
  & Incorrect Constructor Name & Ethereum & VE5 \\
  \cline{2-4}
  & Typographical Error & Ethereum & VE6 \\
  \cline{2-4}
  & Uninitialized Storage Pointer & Ethereum & VE7 \\
  \hline
  \multirow{3}*{Irrelevant Code} & Presence of Unused Variables & Ethereum & VE8 \\
  \cline{2-4}
  & Code With No Effects & Ethereum & VE9 \\
  \cline{2-4}
  & Shadowing State Variables & Ethereum & VE10 \\
  \hline
  \multirow{2}*{Visibility} & Function Default Visibility & Ethereum & VE11 \\
  \cline{2-4}
  & State Variable Default Visibility & Ethereum & VE12 \\
  \hline
  \multirow{3}*{Signature Issue} & Signature Malleability  & Ethereum & VE13\\
  \cline{2-4}
  & Missing Protection against Signature Replay Attacks & Ethereum & VE14 \\
  \cline{2-4}
  & Lack of Proper Signature Verification & Ethereum & VE15 \\
  \hline
  \multirow{3}*{Data Issue} & Unencrypted Private Data On-Chain  & Ethereum & VE16\\
  \cline{2-4}
  & Fake EOS & ESOIO & VS1 \\
  \cline{2-4}
  & Forged Notification, Fake Receipt & ESOIO & VS2 \\
  \hline
  \multirow{3}*{Unprotected Low-level Function} & Unsafe Suicide & Ethereum & VE17\\
  \cline{2-4}
  & Authorization through \textit{tx.origin} & Ethereum & VE18 \\
  \cline{2-4}
  & Unsafe Delegatecall & Ethereum & VE19 \\
  \hline
  \multirow{3}*{Coding Issue} & Unprotected Ether Withdrawal & Ethereum & VE20\\
  \cline{2-4}
  & Write to Arbitrary Storage Location & Ethereum & VE21 \\
  \hline
  \multirow{3}*{Input Issue} & Assert Violation & Ethereum & VE22\\
  \cline{2-4}
  & Requirement Violation & Ethereum & VE23 \\
  \cline{2-4}
  &Wrong Address & Ethereum & VE24 \\
  \hline
  \multirow{3}*{Incorrect Calculation} & Arithmetic Overflow/Underflow & Ethereum, Fabric & VE25, VH1\\
  \cline{2-4}
  & Call-Stack Overflow & Ethereum & VE26 \\
  \cline{2-4}
  &Asset Overflow & EOSIO & VS3 \\
  \hline
  \multirow{3}*{Denial of Service} & DoS with Failed Call & Ethereum, Fabric & VE27\\
  \cline{2-4}
  & Insufficient Gas Griefing & Ethereum & VE28 \\
  \cline{2-4}
  & DoS with Block Gas Limit & Ethereum & VE29 \\
  \hline
  \multirow{4}*{Use of Low-level Function} & Unchecked Send & Ethereum, Fabric & VE30\\
  \cline{2-4}
  & Arbitrary Jump with Function Type Variable & Ethereum & VE31 \\
  \cline{2-4}
  & Hash Collisions & Ethereum & VE32 \\
  \cline{2-4}
  & Message Call with Hardcoded Gas Amount & Ethereum & VE33 \\
  \hline
  \multirow{4}*{Behavioral Workflow} & Reentrancy & Ethereum, Fabric & VE34\\
  \cline{2-4}
  & Arbitrary Jump with Function Type Variable & Ethereum & VE35 \\
  \cline{2-4}
  & Incorrect Inheritance Order & Ethereum & VE36 \\
  \cline{2-4}
  & Infinite Loop & Ethereum & VE37 \\
  \hline
  \multirow{3}*{Consensus Issue} & Transaction Order Dependence & Ethereum, Fabric & VE38\\
  \cline{2-4}
  & Time Manipulation& Ethereum & VE39 \\
  \cline{2-4}
  &Bad Randomness & Ethereum, EOSIO & VE40, VS4 \\
  \hline
  \end{tabular}
  \end{center}
\end{table*}

\begin{table*}[htp]
  \renewcommand\arraystretch{2}
  \scriptsize
  \center
  \caption{A Comparison of State-of-the-art Smart Contracts Analysis Tools}
  \label{vulner_list_second} 
  \resizebox{0.9\textwidth}{!}{ 
  \begin{tabular}{|p{2.2cm}<{\raggedright}|p{1.8cm}<{\raggedright}|p{0.25cm}<{\raggedright}|p{0.25cm}<{\raggedright}|p{0.25cm}<{\raggedright}|p{0.25cm}<{\raggedright}|p{0.25cm}<{\raggedright}|p{0.25cm}<{\raggedright}|p{0.25cm}<{\raggedright}|p{0.25cm}<{\raggedright}|p{0.5cm}<{\raggedright}|l|}\hline
  
    \multicolumn{1}{|c|}{\multirow{2}{*}[-10ex]{\textbf{PROPOSAL}}} & \multicolumn{1}{c|}{\multirow{2}{*}[-10ex]{\textbf{VENUE}}} &\multicolumn{6}{c|}{\textbf{Methodology}} & \multicolumn{2}{c|}{\textbf{Input}}& \multicolumn{1}{c|}{\multirow{2}{*}[-10ex]{\rotatebox{90}{\thead{Open Source}}}}  & \multicolumn{1}{c|}{\multirow{2}{*}[-10ex]{\thead{Vulnerability ID}}}\\ \cline{3-10}
    & & \rotatebox{90}{Formal Verification}  & \rotatebox{90}{\thead{Symbolic Execution}} & \rotatebox{90}{Fuzzing} & \rotatebox{90}{Machine Learning}  & \rotatebox{90}{IR} & \rotatebox{90}{Runtime Verification} & \rotatebox{90}{\thead{Bytecode}}  & \rotatebox{90}{\thead{Source code}} &   & - \\ \hline \hline
    DefectChecker \cite{Chen2022}& IEEE TSE 2022 &  & \checkmark  & & & & &\checkmark & & \cite{DefectChecker2018} & VE27,VE30,VE34,VE38,VE40 \\ \hline
    ReDefender \cite{Pan2021} &IEEE TR 2022 &  & &\checkmark  & & & & &\checkmark &-- &VE34 \\ \hline
    SmartMixModel \cite{Shakya2022} &Blockchain 2022 &  & &\checkmark  & & & &\checkmark &\checkmark &-- &VE34 \\ \hline
    SolSEE \cite{Lin2022} & ESEC/FSE 2022 &  &\checkmark &  & & & & &\checkmark &-- & \\ \hline
    ExGen \cite{jin2022exgen} & IEEE TDSC 2022 &  &\checkmark &  & &\checkmark & & &\checkmark &-- & VE17, VE19, VE25, VH1, VE30, VE34\\ \hline
    solgraph  &-- &  & &  & & & & &\checkmark & \cite{solgraph} & - \\ \hline
    EtherGIS \cite{ZengHZLYTL22} & COMPSAC 2022 & & & &\checkmark & & &\checkmark & &-- & VE17, VE18, VE19, VE34, VE38, VE39\\ \hline
    Vulpedia \cite{YeMLMXZ22} & J.Syst.Softw. & & & &\checkmark & & & & &-- & \\ \hline
    EOSIOAnalyzer \cite{LiHZYLLLTLZ22} & COMPSAC 2022 &  & & & & \checkmark& & \checkmark & & & \\ \hline
    VRust \cite{CuiZGT022} & CCS 2022 &  & & & & \checkmark& & \checkmark & & & \\ \hline
    VetSC \cite{duan2022towards} & CCS 2022 & \checkmark & & & & & & &\checkmark &\cite{VetSC} & \\ \hline
    CodeNet \cite{hwang2022} & IEEE Access 2022 &  & & &\checkmark & & & &\checkmark & & VE18, VE30, VE34, VE39  \\ \hline
    eTainter \cite{ghaleb2022} & ISSTA 2022 &  & & & & & & \checkmark & & & VE28, VE29  \\ \hline
    SmartFast \cite{SmartFast} & Empir.Softw.Eng. &  & & & & \checkmark& &  &\checkmark & & VE2, VE13, VE18, VE19, VE30, VE34, VE37, VE39 \\ \hline
    SolChecker \cite{dong2022solchecker} & CNCIT 2022 & \checkmark & & & & & &  &\checkmark & & VE6,VE7, VE8, VE11 VE18, VE24 VE30, VE34, VE40 \\ \hline
    SVChecker \cite{yuan2022svchecker} & ICCAIS 2022 &  & & & \checkmark & & &  & \checkmark & \cite{SVChecker} & VE18, VE25, VE30, VE34, VE38, VE39 \\ \hline
    SKLEE \cite{jain2022sklee}  & SEFM 2022 &   & \checkmark & & & & &  & \checkmark  & \cite{SKLEE} & VE17, VE25, VE30, VE38, VE39, VE40 \\ \hline
    Sailfish \cite{bose2022sailfish}  & IEEE S\&P 2022 &   &  & & & & \checkmark&  & \checkmark  & \cite{sailfish} & VE34, VE38 \\ \hline
    SciviK \cite{lin2022scivik} & MVSZ & & & & & \checkmark & & &\checkmark &  &  \\ \hline
    EthVer \cite{Mazurek21} & FC 2021 & \checkmark & & & & & & \checkmark & &\cite{ethver} & VE34\\ \hline
    SmartPulse \cite{stephens2021smartpulse}  & IEEE S\&P 2021 &  &  & & & \checkmark & &  & \checkmark & \cite{SmartPulse} & \\ \hline
    Pluto \cite{ma2021pluto}  & IEEE TSE 2021 & & \checkmark  & & & & &  & \checkmark & \cite{Pluto}  & VE25, VE34, VE39\\ \hline
    Horus \cite{ZhouQTLG21} & IEEE S\&P 2021 &  \checkmark & & & & &\checkmark & \checkmark & & \cite{Horus} & VE17,VE19,VE24,VE25,VE30,VE34, \\ \hline
    ConFuzzius \cite{TorresIGS21} & Euro S\&P 2021 &\checkmark  &  & \checkmark & & & & &  \checkmark & \cite{ConFuzzius} & VE17,VE19,VE22,VE34,VE25,VE38,VE40\\ \hline
    Gas Gauge \cite{abs-2112-14771} & arXiv 2021 &  & & \checkmark & &\checkmark & & &\checkmark  &\cite{gasgauge} &VE28,VE29\\ \hline
    Eth2Vec \cite{AshizawaYCO21} & BSCI 2021 &  & & & \checkmark& & & & \checkmark & \cite{eth2vec} &VE12,VE29,VE34,VE25,VE39 \\ \hline
    NeuCheck \cite{LuWZSE21} & SPE 2021 &  & & & &\checkmark & & &\checkmark & \cite{NeuCheck} &VE32,VE34,VE25,VE40 \\ \hline
    SmartScan \cite{SamreenA21}& ICSE 2021 & \checkmark & & & & & &\checkmark & &-- &VE27\\ \hline
    SolGuard \cite{PraitheeshanPZJ21} & Inf. Sci. 2021 & \checkmark & & & & & & \checkmark & & -- &VE19,VE30\\ \hline
    Solidifier \cite{Antonino021} & SAC 2021 & \checkmark & & & & & &\checkmark & &--&VE22 \\ \hline
    ReDetect \cite{YuSYJ21} & MSN 2021 &  & \checkmark & & & & & &\checkmark &--&VE34 \\ \hline
    GasChecker \cite{ChenFLZLLXCZ21} & IEEE TETC 2021 & & & \checkmark & & & & &\checkmark & -- &VE28,VE29\\ \hline
    SmarTest \cite{SoHO21} & USENIX  2021 &  & \checkmark  & & & & & &\checkmark &\cite{VeriSmart} &VE17,VE22,VE25,VE30 \\ \hline
\end{tabular}
}
\end{table*}

\begin{table*}[htp]
  \renewcommand\arraystretch{2}
  \scriptsize
  \center
  \resizebox{0.9\textwidth}{!}{ 
  \begin{tabular}{|p{2cm}<{\raggedright}|p{1.8cm}<{\raggedright}|p{0.25cm}<{\raggedright}|p{0.25cm}<{\raggedright}|p{0.25cm}<{\raggedright}|p{0.25cm}<{\raggedright}|p{0.25cm}<{\raggedright}|p{0.25cm}<{\raggedright}|p{0.25cm}<{\raggedright}|p{0.25cm}<{\raggedright}|p{0.5cm}<{\raggedright}|l|}\hline
    Frontrunner-Jones \cite{TorresCS21,ZhouQTLG21} & USENIX  2021 IEEE S\&P 2021 & \checkmark   &   & & & & &\checkmark & & \cite{Frontrunner-Jones} & VE38 \\ \hline
    ContractWard \cite{WangSXLWS21}  & IEEE TNSE 2021 &  & & &\checkmark  & & & & \checkmark& --&VE19,VE30,VE34,VE25,VE38,VE39\\ \hline
    ESCORT \cite{abs-2103-12607} & arXiv 2021 & & & & \checkmark& & & &\checkmark &-- &VE22,VE17,VE27,VE34,VE26,VE38\\ \hline
    Conkas  & -- & &\checkmark & & &\checkmark & &\checkmark &\checkmark &\cite{conkas} & VE25,VE30,VE34,VE38,VE39 \\ \hline
    VeriSmart \cite{SoLPLO20} & IEEE S\&P 2020 & \checkmark & & & & & & & \checkmark & \cite{VeriSmart} &VE17,VE18,VE22,VE25,VE34 \\ \hline
    VeriSolid \cite{MavridouLSD19} & FC 2019 & \checkmark & & & & & & & \checkmark& \cite{anmavrid} & VE24,VE17,VE19,VE30,VE34\\ \hline
    ESBMC \cite{FrankAH20} & USENIX 2020 & & \checkmark &  & & & & \checkmark &  & \cite{esbmc} & \\ \hline
    sFuzz \cite{NguyenP0L020} & ICSE 2020 & & & \checkmark & &  & & \checkmark & &\cite{sfuzz} & VE19,VE29,VE30,VE34,VE25,VE39,VE40\\ \hline
    GNNSCVD \cite{Zhuang2021} \cite{ZhuangLQLWH20} & IJCAI 2020 & & & & \checkmark & & & &\checkmark  & \cite{GraphDeeSmartContract} & VE34,VE39,V40 \\ \hline
    ContractGuard \cite{WangHXZC20}& IEEE TSC 2020 & & & & &  &\checkmark & \checkmark &   & \cite{contractguard} & VE18,VE19,VE30,VE34,VE25\\ \hline
    Ethainter \cite{BrentGLSS20} &PLDI 2020 & \checkmark & & & & & &\checkmark & &-- & VE17,VE18,VE19\\ \hline
    eThor \cite{SchneidewindGSM20} & CCS 2020 & \checkmark & & & & & &\checkmark & &-- & VE34\\ \hline
    SmartSheild \cite{ZhangMLLNG20} & SANER 2020 & \checkmark & & & & & &\checkmark & &-- &VE10,VE30,VE25 \\ \hline
    VerX \cite{PermenevDTDV20} &IEEE S\&P 2020 & \checkmark & \checkmark & & & & &\checkmark & & \cite{verx-benchmarks} &VE29,VE30 \\ \hline
    GASOL \cite{AlbertCGRR20} & TACAS(2) 2020 &  & \checkmark & & & & &\checkmark & &-- &VE28,VE29 \\ \hline
    RA \cite{ChinenYCO20} & IEEE ICBC 2020 & \checkmark & \checkmark & & & & &\checkmark & &-- & VE34 \\ \hline
    EthPloit \cite{ZhangWLM20} & SANER 2020 &  & & \checkmark & & & & &\checkmark & -- &VE17,VE18,VE19,VE30 \\ \hline
    Harvey \cite{WustholzC20} & ESEC/FSE 2020 & &  & \checkmark & & & & \checkmark& & --&VE22 \\ \hline
    SODA \cite{0002CLLGZLZCHTL20} & NDSS 2020 &  & &  & & &\checkmark &\checkmark&\checkmark & \cite{SODA} &VE17,VE18,VE19,VE30,VE34,VE25 \\ \hline 
    Solythesis \cite{LiCL20} & PLDI 2020 & &  &  & & &\checkmark & &\checkmark & \cite{Leeleo3x}& \\ \hline
    TxSpector \cite{ZhangZZL20} & USENIX  2020 & &  &  & & &\checkmark &\checkmark & & \cite{TxSpector} & VE20,VE17,VE18,VE27,VE30,VE34\\ \hline
    Gastap \cite{AlbertGRS19} & VECoS 2019 & & \checkmark &  & & & &\checkmark & & --& VE28,VE29 \\ \hline
    ÆGIS \cite{TorresBNJ19, TorresBNPJM20}  & ASIA CCS 2020 & & & & & &\checkmark & \checkmark &  &\cite{Aegis} & VE17,VE19,VE34 \\ \hline
    Echidna   & -- & & &\checkmark & & & & & \checkmark  &\cite{echidna} & \\ \hline
    Octopus   & -- & &\checkmark & & &\checkmark & &\checkmark &   &\cite{Octopus} & \\ \hline
    Qian et al. \cite{QianLHZW20} & Access 2020 & &  & &\checkmark & & & \checkmark &  &-- & VE34 \\ \hline
    Sereum \cite{RodlerLKD19} & NDSS 2019 & & \checkmark & & & &\checkmark & \checkmark &  & \cite{eth-reentrancy-attack-patterns} & VE34 \\ \hline
    Annotary \cite{WeissS19} & ESORICS 2019 & &\checkmark & & & &  & \checkmark & &\cite{annotary-sublime-plugin} & VE19,VE30,VE36 \\ \hline
    Mythril  & - & \checkmark & \checkmark & & & & &\checkmark & & \cite{mythril} & VE29 \\ \hline
    SolidityCheck \cite{abs-1911-09425} & arXiv 2019 & \checkmark & & & & & & & & \cite{SolidityCheck} & VE34,VE25 \\ \hline
    Slither \cite{FeistGG19} & ICSE 2019 &  & & & &\checkmark & & & \checkmark & \cite{slither} & VE10,VE20,VE17,VE29 \\ \hline
    solc-verify \cite{HajduJ19} & VSTTE 2019 & \checkmark & & & & & & &\checkmark & \cite{SRI-CSL} & VE29,VE25 \\ \hline
    HoneyBadger \cite{TorresSS19} & USENIX  2019 & & \checkmark & & & & &\checkmark & & \cite{HoneyBadger} & VE10,VE20,VE36,VE25 \\ \hline
    ILF \cite{HeBATV19} & CCS 2019& & \checkmark& \checkmark & \checkmark & & & \checkmark &\checkmark & \cite{ilf} & VE17,VE15,VE19,VE20,VE30,VE39 \\ \hline
    Vultron \cite{WangLLML19} & ICSE 2019 & & & \checkmark & & & & &\checkmark & \cite{vultron} & VE19,VE29,VE30,VE34,VE25 \\ \hline
  \end{tabular}
  }
\end{table*}

\begin{table*}[htp]
  \renewcommand\arraystretch{2}
  \scriptsize
  \center
  \resizebox{0.9\textwidth}{!}{ 
  \begin{tabular}{|p{2.1cm}<{\raggedright}|p{1.8cm}<{\raggedright}|p{0.25cm}<{\raggedright}|p{0.25cm}<{\raggedright}|p{0.25cm}<{\raggedright}|p{0.25cm}<{\raggedright}|p{0.25cm}<{\raggedright}|p{0.25cm}<{\raggedright}|p{0.25cm}<{\raggedright}|p{0.25cm}<{\raggedright}|p{0.5cm}<{\raggedright}|l|}\hline
    SIF \cite{PengAR19} & APSEC 2019 & \checkmark & & & & & &\checkmark &\checkmark &\cite{SIF} &  \\ \hline
    SolAnalyser \cite{AkcaRP19} & APSEC 2019 & \checkmark & & & & & &\checkmark & &\cite{SolAnalyser} & VE18,VE29,VE30,VE25,VE39 \\ \hline
    sCompile \cite{ChangGX00Y19} & ICFEM 2019 & & \checkmark & & & & &\checkmark & & --& VE24,VE17,VE19,VE29,VE30 \\ \hline
    FEther \cite{YangL19} & IEEE Access 7 & \checkmark & \checkmark  & & & & & &\checkmark &-- & \\ \hline
    NPChecker \cite{WangZS19} & OOPSLA 2019 & & & &  & \checkmark & & \checkmark  & &-- &VE30,VE34,VE38,VE39 \\ \hline
    SoliAudit \cite{LiaoTHT19} & IOTSMS 2019 & & & \checkmark & \checkmark&   &  &\checkmark &\checkmark & --& VE24,VE18,VE27,VE34,VE25,VE38,VE40\\ \hline
    Manticore \cite{MossbergMHGGFBD19} & ASE 2019 & & \checkmark & & & & &  & & \cite{manticore} & \\ \hline
    VeriSol \cite{wang2018formal} & ICFEM &  \checkmark & & & & & & &\checkmark & \cite{verisol} & -\\ \hline
    solhint  & - &  & & & & & & &\checkmark & \cite{solhint} &- \\ \hline
    KEVM \cite{HildenbrandtSRZ18} & CSF 2018 & \checkmark & & & & & & \checkmark &  & \cite{evm-semantics} & VE25 \\ \hline
    Isabelle/Hol \cite{AmaniBBS18} & CPP 2018 & \checkmark & & & & & &\checkmark & & \cite{eth-isabelle} & VE25\\ \hline
    EtherTrust  & CAV 2018 &  \checkmark & & & & & &\checkmark & & \cite{rauljordan} & VE34\\ \hline
    SolMet \cite{Hegedus18} & WETSEB 2018 &  \checkmark & & & & & &\checkmark & & \cite{SolMet} & VE34\\ \hline
    Vandal \cite{abs-1809-03981} & arXiv 2018 & & & & & \checkmark & &\checkmark &  & \cite{usyd-blockchain} &VE17,VE18,VE30,VE34 \\ \hline
    Maian \cite{NikolicKSSH18} & ACSAC 2018 & & \checkmark & & & & &\checkmark & & \cite{MAIAN} & VE17,VE20,VE30 \\ \hline
    Securify v2.0 \cite{TsankovDDGBV18} & CCS 2018 & & \checkmark & & & & &\checkmark & & \cite{securify2} & VE29\\ \hline
    teEther \cite{KruppR18} & USENIX  2018 & &  \checkmark & & & & & \checkmark& & \cite{teether} & VE17,VE19,VE27,VE30 \\ \hline
    Osiris \cite{TorresSS18} & ACSAC 2018 & & \checkmark& & & & &\checkmark & & \cite{Osiris} & VE25\\ \hline
    SaferSC \cite{abs-1811-06632} & arXiv 2018 & & & & \checkmark  & & & \checkmark& &\cite{Safe-SmartContracts} & VE17,VE20,VE30\\ \hline
    ContractFuzzer \cite{0001LC18} & ASE 2018 & & & \checkmark &  & & & \checkmark& & \cite{ContractFuzzer} & VE19,VE27,VE29,VE30,VE34,VE39,VE40\\ \hline
    F framework \cite{GrishchenkoMS18} & POST 2018 & \checkmark & & & & & &\checkmark & &-- & VE30,VE34,VE38,VE39,VE40\\ \hline
    Zeus \cite{KalraGDS18} & NDSS 2018 & \checkmark & & & & & & &\checkmark & -- & VE18,VE30,VE34,VE25,VE38\\ \hline
    SASC \cite{ZhouHPSNYK18} & NTMS 2018 & & \checkmark  & & & & & &\checkmark  &-- & VE18,VE26,VE37,VE39 \\ \hline
    Reguard \cite{LiuLCCCR18} & ICSE 2018 & & & \checkmark & & & & & \checkmark &-- & VE34 \\ \hline
    MadMax \cite{GrechKJBSS18} & PACMPL 2018 & &  & & &\checkmark & & \checkmark& & \cite{MadMax} & VE29,VE25 \\ \hline
    ContractLarva \cite{AzzopardiEP18} &  RV 2018 & & & & & & & \checkmark & &\cite{contractLarva} & --\\ \hline
    Smartcheck \cite{TikhomirovVITMA18} & WETSB 2018 & & & & & & &  &\checkmark  & \cite{smartcheck} & VE18,VE27,VE30,VE34,VE25,VE39,V40\\ \hline
    EthIR \cite{AlbertGLRS18} & ATVA 2018 & &  & &  &\checkmark & & \checkmark & &\cite{ethIR}& \\ \hline
    ECFChecker \cite{GrossmanAGMRSZ18} & POPL 2018 & &  & &  & &\checkmark & \checkmark & &\cite{ECFChecker}& \\ \hline
    Oyente \cite{AtzeiBC16} & CCS 2016 & & \checkmark & & & & & \checkmark & &\cite{oyente} & VE34,VE26,VE37,VE39\\ \hline\hline
    DISTRIBUTION &   &  \rotatebox{90}{27.8\%} &  \rotatebox{90}{25.9\%}  &  \rotatebox{90}{13.0\%} &  \rotatebox{90}{11.1\%} &  \rotatebox{90}{13.9\%} &  \rotatebox{90}{8.3\%}&  \rotatebox{90}{53.9\%} &  \rotatebox{90}{46.1\%}&  & \\ \hline
  \end{tabular}
  }

\end{table*}

\end{document}